\documentclass[%
 aip,
 amsmath,amssymb,
preprint,%
]{revtex4-1}

\usepackage{multirow}
 \usepackage{booktabs}

\usepackage{graphicx}
\usepackage{dcolumn}
\usepackage{bm}

\usepackage[utf8]{inputenc}
\usepackage[T1]{fontenc}
\usepackage{mathptmx,soul}

\begin{document}


\title{
Experimental study on the effects of photon-pair temporal correlations in entangled two-photon absorption}

\author{Samuel Corona-Aquino}
\author{Omar Calder\'{o}n-Losada}
\email{omar.calderon@correo.nucleares.unam.mx}
\author{Mayte Y. Li-G\'{o}mez}
\author{H\'{e}ctor Cruz-Ramirez}
\author{Violeta Alvarez-Venicio}
\author{María del Pilar Carre\'{o}n-Castro}
\author{Roberto de J. Le\'{o}n-Montiel}
\email{roberto.leon@nucleares.unam.mx}
\author{Alfred B. U'Ren}

\affiliation{Instituto de Ciencias Nucleares, UNAM. Circuito Exterior s/n, C.U., Coyoacán, C.P. 04510, Ciudad de México, México\\
}

\date{\today}

\begin{abstract}
Entangled two-photon absorption (ETPA) has recently become a topic of lively debate, mainly due to the apparent inconsistencies in the experimentally-reported ETPA cross sections of organic molecules. In this work, we provide a thorough experimental study of ETPA in the organic molecules Rhodamine B (RhB) and Zinc Tetraphenylporphirin (ZnTPP). The goal of this contribution is twofold: on one hand, it seeks to reproduce the results of previous experimental reports and, on the other, it aims to determine the effects of different temporal correlations ---introduced as a controllable time-delay between the photons to be absorbed--- on the strength of the ETPA signal. In our experiment, the samples are excited by entangled pairs produced by type-I SPDC, with a spectral distribution centered at 810 nm. Surprisingly, and contrary to what was expected, the time delay did not produce in our experiments any systematic change in the cross-sections when monitoring the ETPA signal using a transmission measurement scheme. As a plausible cause of this unexpected result, we argue that the photon-pair flux, typically-used in these experiments, is not sufficient to promote the two-photon absorption process in these molecules. This suggests that the actual absorption cross-section values are lower than those previously reported, and therefore do not lead to a measurable ETPA effect for the transmission method.
\end{abstract}

\maketitle

\section{Introduction}
Since its inception in 1931, the work of Maria Goeppert Mayer on two-photon absorption (TPA) has become a fundamental tool for probing processes that require a sharp resolution and high energy density concentrations, as in the case of biological imaging and microscopy \cite{denk1990}.
Remarkably, TPA has also improved spectroscopic methods by enabling the study of the symmetry of excited states in organic molecules, including electronic transitions of a system that are normally inaccessible by one-photon absorption (1PA).~\cite{xu_webb_1996,Rumi-2010} 

Despite its broad applicability, TPA is an extremely inefficient process, often requiring large photon fluxes on the order of $10^{27}$ photons s$^{-1}$cm$^{-2}$, which corresponds to an energy of 1.25 nJ per pulse in a femto-second laser centered at 800 nm (100 mW average power)~\cite{Rumi-2010}. This high peak-excitation is required to compensate for the low probability that two photons will almost simultaneously reach a molecule and secure the excitation. Nevertheless, theoretical research developed in the last three decades has posed a promising solution by proposing the use of non-classical light sources so as to induce the two-photon excitation at considerably lower photon fluxes. \cite{Javanainen-1990,Fei-1997,Lee-2006,roslyak2009-1,roslyak2009-2,guzman2010,raymer2013,schlawin2013,schlawin2016} Recent investigations exploit nonlinear processes such as spontaneous parametric down conversion (SPDC) 
and spontaneous four wave mixing (SFWM) 
in order to generate entangled photon pairs to be used as source for TPA in a phenomenon referred to as Entangled Two-Photon Absorption (ETPA). \cite{Saleh-1998,kojima2004,perina1998,DLMontiel-2013,dorfman2016,schlawin2017,oka2010,Schlawin-2018,villabona_calderon_2017,Varnavski2017,oka2018-1,oka2018-2,svozilik2018-1,svozilik2018-2,burdick2018,montiel2019} By virtue of the strong correlations between the photons in each pair, the phenomenon has been reported at significantly lower photon fluxes, on the order of $10^{12}$ photons s$^{-1}$ cm$^{-2}$, which corresponds to a photon rate of $10^6$ pairs/s,~\cite{Harpham-2009} representing a clear advantage over classical TPA. However, a detailed analysis is needed in view of recent findings on the improvement achieved on the fluorescence signal using squeezed light over a coherent source in a two-photon absorption experiment,~\cite{li_SL_2020} where the authors report an enhancement lower than $10^{2}$. This result is in the same direction as a recent work by Raymer \textit{et. al}~\cite{raymer2020Theory} in which a thorough analysis of the quantum advantage is carried out for realistic experimental scenarios, showing that this enhancement should be imperceptible. Note that if a quantum-enabled advantage of ETPA over classical TPA were to be experimentally confirmed, the resulting low photon flux regime would open up the possibility of performing two-photon spectroscopy with more compact and cost-effective devices.

The absorption cross-sections $\delta_\mathrm{TPA}$ 
and $\sigma_\mathrm{E}$ constitute useful metrics  for TPA and ETPA, respectively; the first has typical values on the order of  $\delta_\mathrm{TPA} \sim 10^{-48}$ cm$^{4}$s/photon for most organic molecules.~\cite{upton_2013}  Contemporary investigations have benefited from the expected quantum advantage in the ETPA process leading to a number of experiments on organic molecules including biological samples, such as proteins,~\cite{villabona_2018} thienoacenes,~\cite{eshun_goodson_2018} and commercially available dyes such as fluorescein,~\cite{li_SL_2020} Zinc Tetraphenylporphyrin (ZnTPP) and Rhodamine B (RhB).~\cite{villabona_calderon_2017} These experiments report ETPA cross sections on the order of $\sigma_\mathrm{E} \sim 10^{-18} - 10^{-21}$ cm$^{2}$/molecule.

An important property of ETPA is its dependence on the temporal delay between the absorbed entangled photons. This dependence is such that the absorption is expected to be suppressed if the photons arrive at the molecule with a temporal delay much larger than the lifetime of the intermediate level in question.~\cite{Saleh-1998,perina1998,DLMontiel-2013} A recent report has shown a possible time dependence on the ETPA signal for Rh6G in ethanol.~\cite{tabakaev_2019} However, it is worth pointing out that, although the presented behavior is broadly consistent with theoretical predictions, the interferometric scheme used to control the temporal delay introduces a dependence of the incoming photon-pair flux on the delay, i.e. unrelated to the ETPA process, which ultimately influences the ETPA signal. This possible artifact deserves, without any doubt, a more detailed analysis; particularly in view of a recent work by Landes \emph{et al.,}~\cite{landes2020} in which the authors report no measurable ETPA signal using the same molecular system, and a similar experimental arrangement.

Other related works studying the temporal delay dependence of the ETPA signal have focused on its non-monotonic behavior ~\cite{Lee-2006,Varnavski2017,Varnavski2020} rather than its suppression outside the time-correlation of the absorbed photons. A recent work by Mikhaylov \emph{et al}.~\cite{Mikhaylov-2020} has drawn attention to the difficulties in discriminating the effects due to ETPA from those resulting from other (linear) phenomena which are not dependent on the temporal delay. Because in that work the introduction of a deterministic delay is not possible, i.e. part of the photon-pair flux always reaches the sample without delay,\cite{raymer2018} full ETPA suppression may not be observed directly. One of the main motivations of our work is to perform an ETPA experiment with a deterministic temporal delay introduced between the two photons in each pair. Along this line, the goal of this paper is to assess the viability of the transmission-based scheme, commonly-used to measure the ETPA cross-section values of organic molecules. The first part of our experiment tests the conclusions of previous experimental reports and compares them to ours. The second part of the experiment introduces a deterministic intra-pair temporal delay as an additional parameter to examine the validity of such conclusions. In addition, we present a new model that accounts for time-independent effects and therefore effectively estimates the ETPA cross-sections. Conveniently, we have selected the organic molecules RhB and ZnTPP for our ETPA experiments so as to facilitate comparisons with previous works.~\cite{upton_2013,villabona_calderon_2017,Mikhaylov-2020}

\section{Theory}

It has been shown that down-converted light exhibits a high degree of correlation so that the detection of one of the photons in a given pair, at a particular space-time location, determines the corresponding location where the detection of its twin is expected.~\cite{Shih-2003} Due to the fact that in a photon-pair stream the presence of any one photon is accompanied by its twin, the probability of excitation of an atom (or molecule) by an individual photon pair is independent of the flux and consequently the two-photon absorption process exhibits a transition rate with a linear dependence on the incident flux~\cite{Javanainen-1990}.

Let us consider the total number of excited molecules (per unit volume) during an interaction time, $\tau$,
\begin{equation}
    n_\mathrm{molec}^{(2)} \approx \frac{1}{2} \sigma_E\phi \tau n_0,
\end{equation}
where $\sigma_E$ denotes the entangled-photon absorption cross-section of a molecule, $\phi=N_\mathrm{ph}/(A\tau)$ is the photon flux (with $N_\mathrm{ph}$ the number of incident photons and $A$ the transverse area of the photon-pair spatial mode),  and $n_0 = c N_A$ is the density of molecules (with $c$ the concentration and $N_A$ Avogadro's number). As a result, the total number of absorbed biphotons,  $N_\mathrm{ph}^{(2)}$, in an interaction volume $V$, is expressed as follows
\begin{equation}
    N_\mathrm{ph}^{(2)} \approx 2 n_\mathrm{molec}^{(2)} V = \frac{\sigma_E c N_A V}{A} N_\mathrm{ph}.
\end{equation}
Interestingly, from the experimental parameters, we can obtain the following expression
\begin{equation}
\sigma_{_{E}}=\frac{\left(N_\mathrm{ph}^{(2)}/N_\mathrm{ph}\right)A}{cVN_{A}}=\frac{m}{c\ell N_{A}},
\label{eq:sigmastandar}
\end{equation}
where $m$ is the slope calculated from the linear relationship between $N_\mathrm{ph}^{(2)}$ and $N_\mathrm{ph}$ at different concentrations. Note that the right-most expression in equation~(\ref{eq:sigmastandar}) is obtained by approximating the interaction volume, determined by the
photon-pair spatial mode, as a cylinder with transverse area $A$ and length  $\ell \leq z_{R}$, where $\ell$ is the length of the cuvette which contains the molecules and $z_{R}$ is the Rayleigh range. For the specific case of a cuvette of $\ell=1$ cm length,  this approximation is valid if the biphoton spatial mode is focused to a spot size with radius $\sim 36$ $\mu$m.

. 

\subsubsection*{Sensitivity of the transmittance scheme}

In typical transmission-based ETPA measurements, two different signals are needed: a reference signal, which is obtained by probing a cuvette containing the solvent only ($R_\mathrm{solv}$), and a second signal, obtained with a solution of the sample under study ($R_\mathrm{sam}$). The difference between these signals yields the absorption rate ($R_\mathrm{abs}$). Because the reference-signal values take into account linear effects produced by the solvent and the cuvette, $R_\mathrm{abs}$ arguably exhibits only the effects due to the ETPA process.



In what follows we describe a method for estimating the sensitivity of an instrument which relies on transmission measurements for obtaining ETPA cross-sections. Note that because our desired measurement is based on photon counting, the uncertainty of the measured rates can be estimated by the square root of their mean value. Therefore, a statistically significant difference (between the signals) will occur if
\begin{equation}
    \bar{R}_\mathrm{sam} + \sqrt{\bar{R}_\mathrm{sam}}/2 \leq \bar{R}_\mathrm{solv} - \sqrt{\bar{R}_\mathrm{solv}}/2,
\end{equation}
where the over-bar symbol represents the mean value. As a consequence, the quantity $b\equiv 1/\sqrt{\bar{R}_{solv}}$ serves as a lower bound for the ratio $\bar{R}_\mathrm{abs}/\bar{R}_\mathrm{solv}$,
\begin{equation}
   \frac{\bar{R}_\mathrm{abs}}{\bar{R}_\mathrm{solv}} \geq \frac{1}{\sqrt{\bar{R}_\mathrm{solv}}}\equiv b.
\end{equation}
Using the definition of the cross-section in equation~(\ref{eq:sigmastandar}), $\sigma_E$ will be bounded by $\sigma_E^{\mathrm{LB}}$, i.e. $\sigma_E \ge \sigma_E^{\mathrm{LB}}$, with
\begin{equation}\label{eq:lowerbound}
    \sigma_E^\mathrm{LB}=\frac{b}{c \ell N_\mathrm{A}}.
\end{equation}

In a typical experiment using SPDC light,~\cite{Kurtsiefer-2001,Suzer-08} $\bar{R}_\mathrm{solv} \sim 10^5 $ photons/s, which implies that the bound is $b \sim 10^{-3}$, this means that for concentrations in the range of tens of $\mu$M, the lower bound is on the order of ${\sigma}_E^\mathrm{LB} \sim 10^{-19}$ cm$^2$/molecule, whereas for concentrations in the range of tens of mM the lower bound is on the order of $10^{-22}$ cm$^2$/molecule.

\subsubsection*{Biphoton absorption rates}

In order to obtain a value for the ETPA signal $\bar{R}_\mathrm{abs}/\bar{R}_\mathrm{solv}=1-\bar{R}_\mathrm{samp}/\bar{R}_\mathrm{solv}$,  previous works rely on directly computing the quotient $\bar{R}_\mathrm{samp}/\bar{R}_\mathrm{solv}$ from the singles or coincidence counts obtained in transmission through the cuvette, containing the sample $\bar{R}_\mathrm{samp}$ and the solvent only $\bar{R}_\mathrm{solv}$.  This approach, however, can be misleading, since it ignores the various losses throughout the experiment. Therefore, we follow  Schneeloch {\it et al.,}~\cite{Schneeloch-2019} using a simple model which relies on the actual photon-pair flux
expected at the cuvette.  Let us assume that $R^{(2)}$ biphotons per second reach the cuvette, and after the ETPA sample are separated into two arms by a 50/50 beamsplitter. The resulting photons in each of the two arms are  coupled into an optical fiber and detected at a single photon detector. This results in a rate of singles detection for each of the two channels ($R_1$, $R_2$) and a coincidence rate $R_{12}$, expressed as follows
\begin{subequations}\label{eq:measrates}
\begin{align}
    R_1 & = (\epsilon_1 \kappa_1 \beta_1) R^{(2)} + \varphi_1,\\
    R_2 & = (\epsilon_2 \kappa_2 \beta_2) R^{(2)} + \varphi_2,\\
    R_{12} & = (\epsilon_1\epsilon_2 \kappa_1 \kappa_2 \beta_{12}) R^{(2)} + \varphi_{12},
\end{align}
\end{subequations}
with the coupling efficiencies being $\kappa_1$ and $\kappa_2$ for the signal and idler arms, respectively. The linear absorption, detector efficiencies, scattering and all other losses are taken into account in the transmission efficiencies $\epsilon_1$ and $\epsilon_2$, while the probabilities that the photons in each pair are transmitted, reflected or separated by the beamsplitter are described by $\beta_1$, $\beta_2$ and $\beta_{12}$, respectively (with values $\beta_1=\beta_2=3/4$, and $\beta_{12}=1/2$).  We also incorporate in this simple model the dark-count rate in each of the two detectors ($\varphi_1$ and $\varphi_2$), as well as accidental coincidence count rate $\varphi_{12}$. Note that by combining the set of equations in~(\ref{eq:measrates}), the biphoton rate which reaches the cuvette can be expressed as
\begin{equation}\label{eq:measbirate}
    R^{(2)} = \left(\frac{\beta_{12}}{\beta_1\beta_2}\right)\frac{\tilde{R}_1\tilde{R}_2}{\tilde{R}_{12}},
\end{equation}
where the symbols with a tilde represent each of the rates corrected for dark or accidental counts, i.e,  $\tilde{R}_\mu=R^{(2)}-\varphi_\mu$, with $\mu \in \{1,2,12\}$.   Interestingly, because of the structure of this last expression, we are able to correctly estimate the two photon flux $R^{(2)}$, irrespective of linear losses quantified by $\epsilon_{1}$, $\epsilon_{2}$,$\kappa_{1}$, and $\kappa_{2}$. Note from Equation~(\ref{eq:measrates}), that the effect of ETPA is to reduce the photon-pair flux from $R^{(2)}_\mathrm{solv}$ to  $R^{(2)}_\mathrm{samp}$, according to $R^{(2)}_\mathrm{samp} =  \epsilon_\mathrm{ETPA} R^{(2)}_\mathrm{solv}$,
where $\epsilon_\mathrm{ETPA}$ represents the photon-pair flux reduction produced by ETPA.

We can now use Equation~(\ref{eq:measbirate}), considering that $\beta_1$, $\beta_2$, and $\beta_{12}$ are all independent of whether or not ETPA takes place, to express the ratio of absorbed to incident biphotons as
\begin{equation}
\label{eq:biphotonratios}
    \frac{\bar{R}^{(2)}_\mathrm{abs}}{\bar{R}^{(2)}_\mathrm{solv}} = 1 - \frac{\bar{R}^{(2)}_\mathrm{samp}}{\bar{R}^{(2)}_\mathrm{solv}} = 1- \epsilon_\mathrm{ETPA}=
    1 -   \frac{\tilde{R}_1^\mathrm{samp}\tilde{R}_2^\mathrm{samp}/\tilde{R}_{12}^\mathrm{samp}}{\tilde{R}_1^\mathrm{solv}\tilde{R}_2^\mathrm{solv}/\tilde{R}_{12}^\mathrm{solv}} = 1 -  \frac{g^{(2)}_{0,\mathrm{solv}}}{g^{(2)}_{0,\mathrm{sam}}},
\end{equation}
where we have introduced the second order correlation function $g^{(2)}_{0} = \tilde{R}_{12}/(\tau_c \tilde{R}_{1}\tilde{R}_{2})$, written in terms of the coincidence window  $\tau_c$. From Equations (\ref{eq:sigmastandar}) and (\ref{eq:biphotonratios}), we may then express the cross-section as follows
\begin{equation}
\label{eq:sigma_g2}
    \sigma_E=\frac{1 -  g^{(2)}_{0,\mathrm{solv}}/g^{(2)}_{0,\mathrm{sam}}}{c \ell N_\mathrm{A}}.
\end{equation}
Remarkably, as is clear from Equation~(\ref{eq:sigma_g2}), we can write the ETPA cross-section in terms of the   second-order correlation functions $g^{(2)}_{0,\mathrm{solv}}$ and $g^{(2)}_{0,\mathrm{sam}}$, which can be obtained experimentally.


In the experimental section below we discuss two experiments, the second of which involves the introduction of a temporal delay $\tau$ between the signal and idler photons prior to reaching the ETPA sample. Because of the specific method used in order to introduce this delay (based on Hong Ou Mandel interference), the available photon-pair flux $R^{(2)}$ depends on $\tau$ (specifically, with $R^{(2)}$ at $\tau=0$ being twice as large as compared to the corresponding value at a large positive or negative delay, see Appendix  \ref{App:ControllingDelay}).  However, considering that the quantities $\epsilon_{1}$, $\epsilon_{2}$, $\kappa_{1}$, $\kappa_{2}$,  $\beta_1$, $\beta_2$ and $\beta_{12}$ are all delay-independent, any delay dependence of the flux, i.e. $R^{(2)}=R^{(2)}(\tau)$,
will cancel out in Eq.~(\ref{eq:biphotonratios}), thus not affecting the ETPA signal $\bar{R}^{(2)}_\mathrm{abs}/\bar{R}^{(2)}_\mathrm{solv}$. Thus, remarkably, we can use Equation~(\ref{eq:sigma_g2}) to monitor the ETPA signal behavior in the transmission scheme, unaffected by any intrinsic dependence of the incoming photon-pair flux on the delay between the absorbed photons, and/or on any linear loss mechanisms sustained by the photon pairs .

Also note that due to the existing linear relationship between the rates (singles and in coincidence) for solvent and samples, i.e., $R_\mu^\mathrm{samp}=m_\mu R_\mu^\mathrm{sol}$ (with $\mu=1,2,12$), the biphoton ratio in Equation~(\ref{eq:biphotonratios}) can be calculated using the expression
\begin{equation}
\label{eq:biphoratio_slopes}
    \frac{\bar{R}^{(2)}_\mathrm{abs}}{\bar{R}^{(2)}_\mathrm{solv}} = 1 - \frac{m_1 m_2}{m_{12}},
\end{equation}
where each $m_\mu$  can be obtained by linear fitting of the corresponding experimental transmission data.

\section{Experimental Configurations and Results}
In order to investigate the ETPA process, we have measured the absorption in RhB and ZnTPP using a transmission scheme based on single-channel and coincidence photon counting. So as to place our current work in the context of previous results, we have calculated the cross-sections using the standard prescription shown in equation~(\ref{eq:sigmastandar}) and compared it to our proposed approach, which makes use of the second-order correlation function in equation~(\ref{eq:sigma_g2}).

As explained above, our experiment is divided in two stages: the first employs a collinear SPDC source, and the second makes use of a non-collinear source. In both configurations, entangled pairs of photons were generated through spontaneous parametric down conversion (SPDC) to co-linearly excite the molecules. Succinctly, SPDC is a nonlinear process in which single photons from an intense laser, interacting with a nonlinear crystal, decay into photon pairs so that both linear momentum and energy are conserved.  In our work, we employ a continuous wave (CW) laser, centered at 405 nm, to pump a $\beta$-Barium Borate  (BBO) crystal which is configured to produce degenerate-type-I SPDC  photons around 810 nm, in both the collinear and non collinear geometries. We have chosen to use type-I SPDC because it yields a larger bandwidth of entangled photons (while permitting, in a non-collinear configuration, the introduction of  a deterministic delay between the signal and idler photons).

In the collinear configuration, we have two degrees of freedom at our disposal:  control of the pump power and the molecular concentration. In the non-collinear configuration, we introduce, as an additional variable, the signal-idler temporal delay.
This allows us to study the dependence of the ETPA signal on the photon-pair coherence (correlation) time.

\subsubsection*{Collinear SPDC configuration}

\begin{figure}[htbp]
    \centering
    \includegraphics[width=0.95\textwidth]{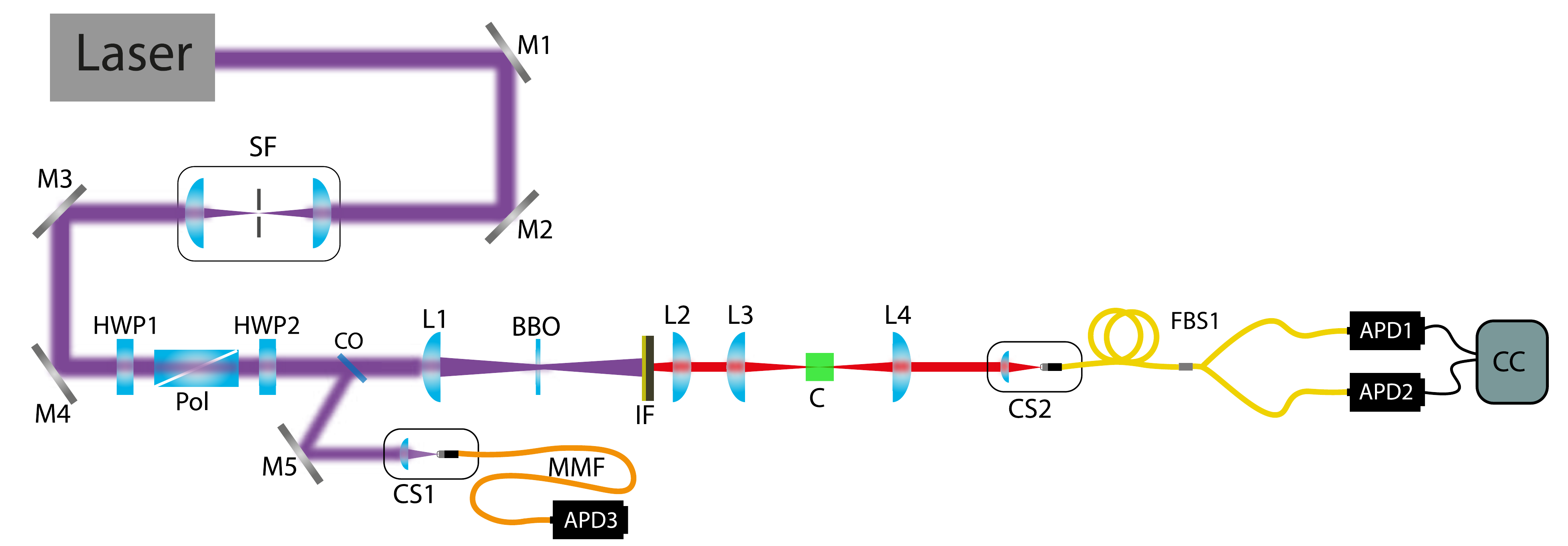}
    \caption{
    Experimental setup used to sudy ETPA with photon pairs generated through collienar type-I SPDC. A spatially-filtered (SF) beam from a CW laser at $\lambda = 404.7$ nm with  controllable power and polarization state (HWP, Pol) pumps a BBO crystal so as to generate SPDC photon pairs centered at 809 nm,  filtered (IF) from the remaining pump, and spatially shaped using plano-convex lenses (L). The pairs interact with a quartz cuvette (C) containing the molecular samples, and the transmitted biphotons are split (FBS) and detected in two avalanche photo-diodes (APD), obtaining singles and coincidence counts, using a time to digital converter (CC).
    }
    \label{fig:colineal}
\end{figure}

The setup for the first experimental configuration is shown in Figure~\ref{fig:colineal}. A CW laser (\textit{MogLabs ECD004}), centered at a wavelength 404.87 nm with a nominal bandwidth of $<300$ kHz, is spatially filtered (SF) using a pinhole preceded and followed by aspheric lenses, resulting in a Gaussian beam. The optical power in the beam is controlled using a rotating half-waveplate (HWP1) followed by a linear polarizer (Pol). A second half-waveplate (HWP2) is introduced so as to match the pump polarization to the crystal orientation and optimize the SPDC generation rate. The resulting pump beam is then focused with a plano-convex lens (L1) of 150-mm focal length on a  $\beta$-barium borate (BBO) crystal of 1mm thickness (\textit{NewLight Photonics}) to generate type-I SPDC.  A weak  reflection from a coverslip (CO) located between HWP2 and L1 is coupled into a multimode optical fiber (MMF) using the coupling system CS1, with the other end of the fiber connected to an avalanche photo-diode (APD3) (\textit{Perkin Elmer-SPCM-AQR-14-FC}) so as to monitor the pump power.

In order to block the remaining pump photons, two interference filters (IF) were placed at the output of the BBO crystal. The first filter is a long-pass filter (\textit{Semrock}), with $>98\%$ transmittance for wavelengths greater than 488 nm, and attenuation in six units of optical density for the rest of the spectrum. It is followed by a bandpass filter centered at 809 nm, with a bandwidth of 81 nm. Once filtered, the photon-pair spatial mode is collimated with a plano-convex lens of focal length 30 mm (L2). It is worth pointing out that without the interference filters, the spectral distribution of the SPDC photons is centered at 809 nm, with bandwidth of $\sim200$ nm. The photon pairs are subsequently focused using a lens (L3) of focal length $f=75$ mm, yielding a spot size $w_{0}$ of $\sim 88$ $\mu$m. At the focal plane, we placed a quartz cuvette (C) (\textit{Hellman Analytics}), which contains the solutions of organic molecules. The cuvette has a 1 cm width, with a transmission of $> 88\%$ in the UV-Visible range.

The spatial mode which exits the cuvette is collimated with a lens (L4) of focal length $f=50$ mm, and coupled, with the help of the coupling system (CS2) into an input port of a single-mode fiber beamsplitter, labeled as FBS1 (\textit{Thorlabs}, TW805R5F2).  The output ports of the fiber beamsplitter are connected to  two avalanche photo-diodes APD1 and APD2, with their electronic signals monitored by a time to digital converter (\emph{IDQuantique ID 800}) (CC), with a coincidence window of 1.05 ns.
We are then able to measure the singles and coincidence count rates,  in order to extract information about the ETPA process.

The molecular samples used in our ETPA experiments were HPLC-rated Rhodamine B (RhB, $\geq$ 95\%) dissolved in methanol and Zinc tetraphenylporphyrin (ZnTPP, $\geq$ 99\%) dissolved in toluene (both chemical compounds from \textit{Sigma Aldrich}).
The singles and coincidence counts are recorded in different experimental configurations, i.e. with the photon-pair stream traversing the quartz cuvette containing, at first, the solvents only (methanol or toluene) and, subsequently, the solutions of RhB or ZnTPP (in methanol and toluene, respectively).  These signals are recorded for different pump power levels, as controlled by a half waveplate HWP1, in such a way that the experimental conditions do not change between solvent-only and solution experiments.

\subsubsection*{Results: Collinear SPDC configuration}

As explained above, in this configuration we consider only two degrees of freedom, namely the pump power and the sample concentration. Note that by varying the pump power, we can control the SPDC photon-pair flux arriving at the molecular samples. While previous works using similar conditions have been reported,~\cite{villabona_calderon_2017} we take advantage of the convenience and simplicity of this configuration to perform absorption tests and compare the cross-sections obtained: i) through what we refer to as the standard scheme, i.e. based on the slope of the absorption rate vs incident photon pair flux dependence, and ii) via our proposed model which makes use of the second-order correlation functions in equation~(\ref{eq:sigma_g2}).

\begin{figure}[htbp]
	\centering
	\includegraphics[width=0.48\textwidth]{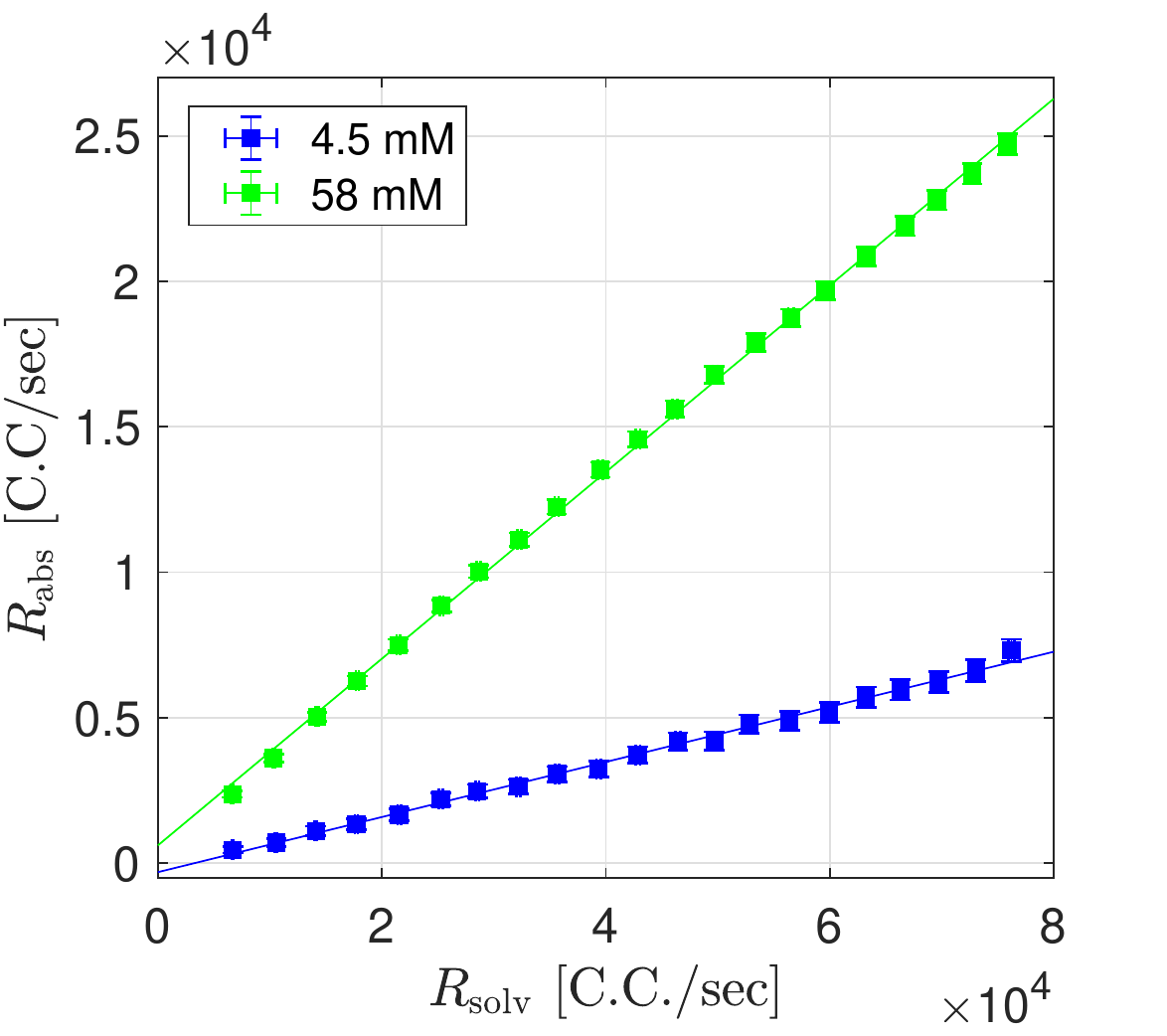}
	\includegraphics[width=0.48\textwidth]{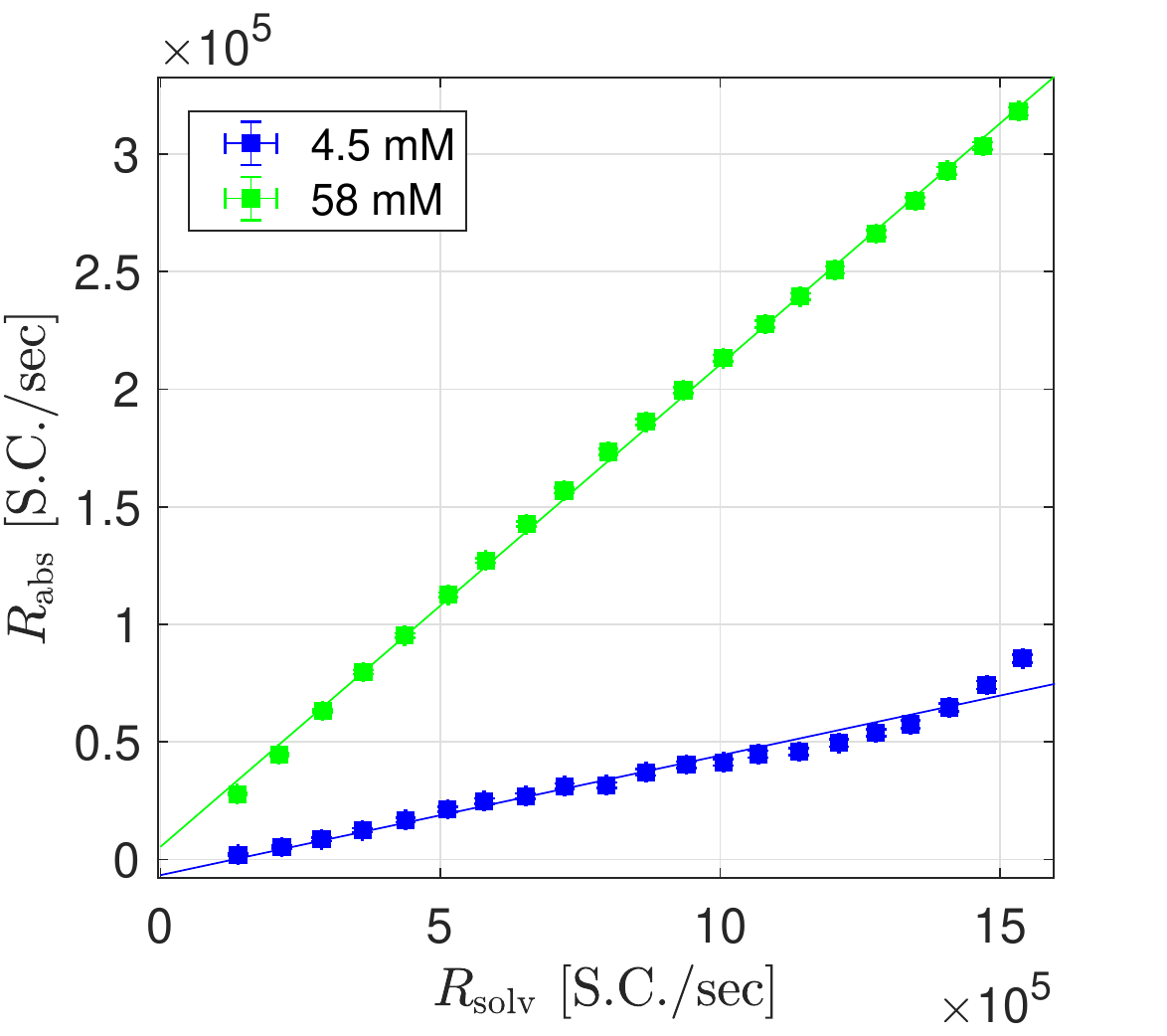}
	\includegraphics[width=0.49\textwidth]{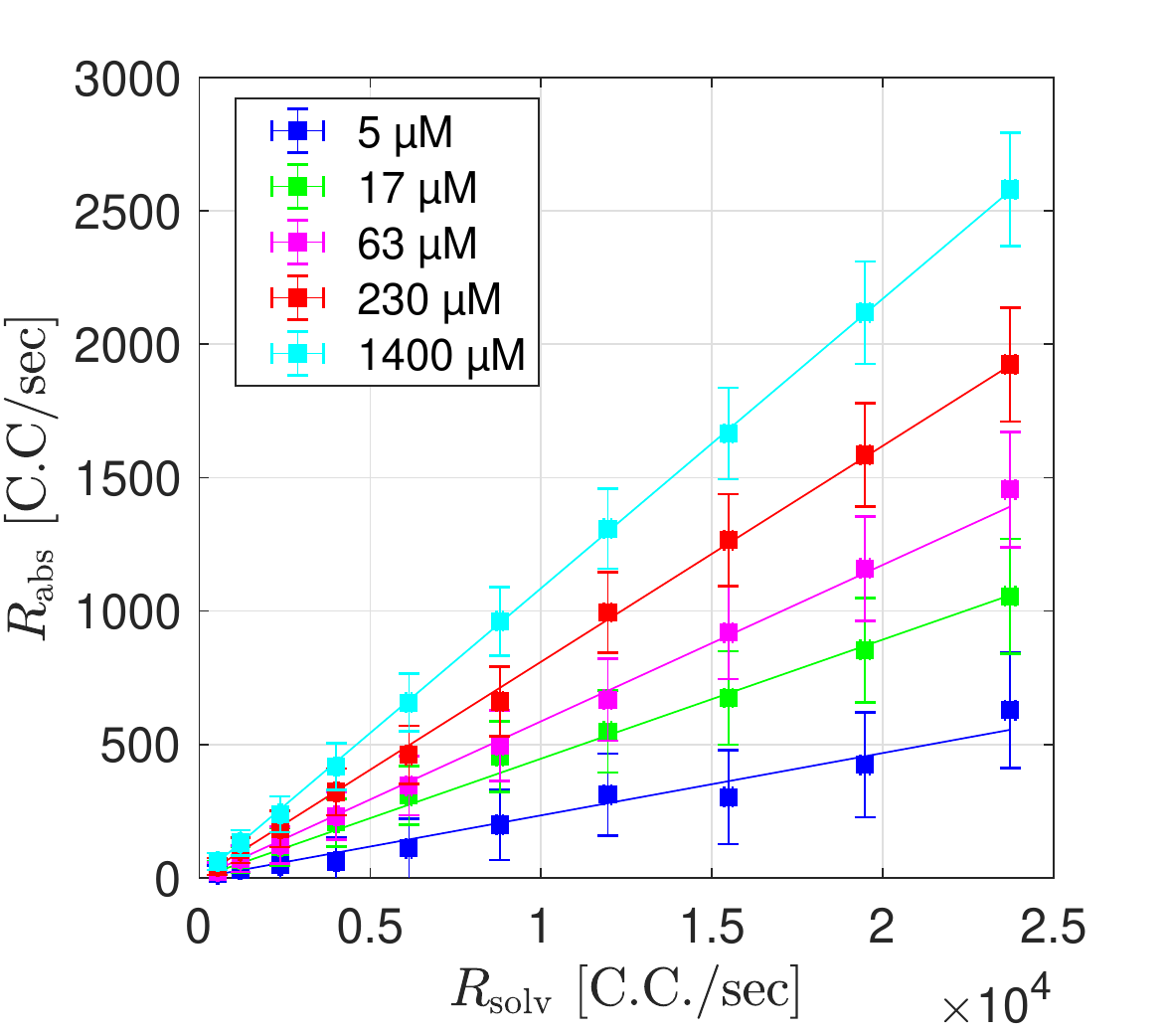}
	\includegraphics[width=0.49\textwidth]{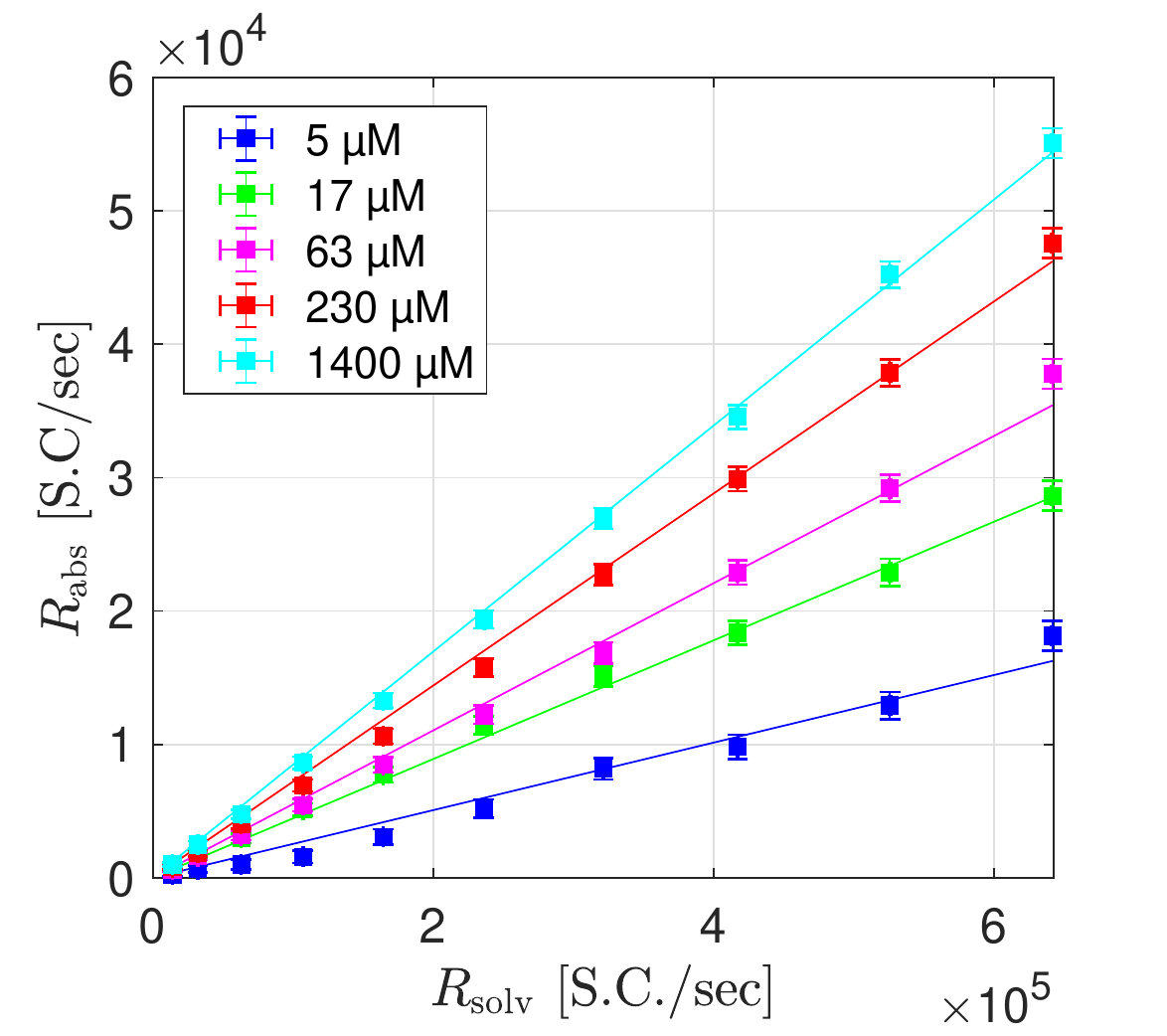}
	\caption{{ETPA} rate measured for RhB (upper row) and ZnTPP (lower row) in coincidence (left column) and single counts (right column) after transmission measurements done in the collinear scheme.}
	\label{fig:AbsRhBZnTppColineal}
\end{figure}

Because our experiment is performed in a low-pump-power regime, a linear dependence of the absorption rate on the incident photon-pair flux is expected for the standard scheme. Indeed, we have observed such a behavior, both for coincidence and singles rates, as shown in Figure~\ref{fig:AbsRhBZnTppColineal}. The figure shows measurements for different concentration levels, as indicated, for RhB in the upper row and for ZnTPP in the lower row, with the coincidence count rate plotted in the first column and the singles count rate in the second column.  The solid dots correspond to the experimental data, with error bars depicting the Poissonian uncertainty, while the solid lines show their respective linear fits.

Making use of equation~(\ref{eq:sigmastandar}) and evaluating the corresponding slope from the linear fit, we have calculated the ETPA cross-section for both molecules. The values that we obtained are reported in the first and second columns of Table~\ref{Table: RhBCol} for RhB and Table~\ref{Table: ZnTPPCol} for ZnTPP, respectively. Note that these cross-sections are in good agreement with the values reported by previous authors.~\cite{villabona_calderon_2017} 

\begin{table}[htp]
	\centering
	\begin{tabular}{>{\centering\arraybackslash}m{1.5cm} >{\centering\arraybackslash}m{3.5cm}
	>{\centering\arraybackslash}m{3.5cm}
	>{\centering\arraybackslash}m{3.5cm}}
	\hline
    \multirow{2}{*}{c (mM)} & \multicolumn{3}{c}{$\sigma_{E}$ [cm$^{2}$/molecule]} \\ \cline{2-4}
     & Eq.~(\ref{eq:sigmastandar}) with C.C &  Eq.~(\ref{eq:sigmastandar}) with S.C & Eq.~(\ref{eq:sigma_g2}) \\
    \hline
	\hline
	4.5	& $3.27(0.18) \times 10^{-20}$ & $1.64(0.19)\times 10^{-20}$ & $0.19(0.14) \times 10^{-20}$ \\
    58	& $9.51(0.13)\times 10^{-21}$ & $6.02(0.18) \times 10^{-21}$ & $2.03(0.23)\times 10^{-21}$  \\
\hline
\end{tabular}
	\caption{ETPA cross-section values,  $\sigma_{E}$, for two different concentrations of {RhB} calculated from different definitions given by the standard prescription (Eq.~(\ref{eq:sigmastandar})) using coincidence (C.C.) and single counts (S.C.), and the second order correlation functions (Eq.~(\ref{eq:sigma_g2})). The number in brackets represents he absolute error.}
\label{Table: RhBCol}
 \end{table}

\begin{table}[htp]
	\centering
	\begin{tabular}{>{\centering\arraybackslash}m{1.5cm} >{\centering\arraybackslash}m{3.5cm}
	>{\centering\arraybackslash}m{3.5cm}
	>{\centering\arraybackslash}m{3.5cm}}
	\hline
    \multirow{2}{*}{c ($\mu$M)    } & \multicolumn{3}{c}{$\sigma_{E}\times 10^{-18}$ [cm$^{2}$/molecule]} \\ \cline{2-4}
     & Eq.~(\ref{eq:sigmastandar}) with C.C &  Eq.~(\ref{eq:sigmastandar}) with S.C & Eq.~(\ref{eq:sigma_g2}) \\
    \hline
	\hline
5	& $8.36(1.06)$ & $7.77(0.70)$ & 9.94(1.23) \\
17	& $4.33(0.22)$ & $4.36(0.38)$ & 4.30(0.44)  \\
63	& $1.45(0.10)$ & $1.54(0.04)$  & 1.40(0.15) \\
230	& $0.519(0.023)$  & $0.584(0.012)$ & 0.474(0.049) \\
1400& $0.100(0.002)$ & $0.128(0.002)$ & 0.074(0.007) \\
\hline
\end{tabular}
	\caption{ETPA cross-section values,  $\sigma_{E}$, for five different concentrations of {ZnTPP} calculated from different definitions given by the standard prescription (Eq.~(\ref{eq:sigmastandar})) using coincidence (C.C.), single counts (S.C.), and the second order correlation functions (Eq.~(\ref{eq:sigma_g2})). The number in brackets represents the absolute error.}
	\label{Table: ZnTPPCol}
 \end{table}

\begin{figure}[htbp]
    \centering
    \includegraphics[width=0.49\textwidth]{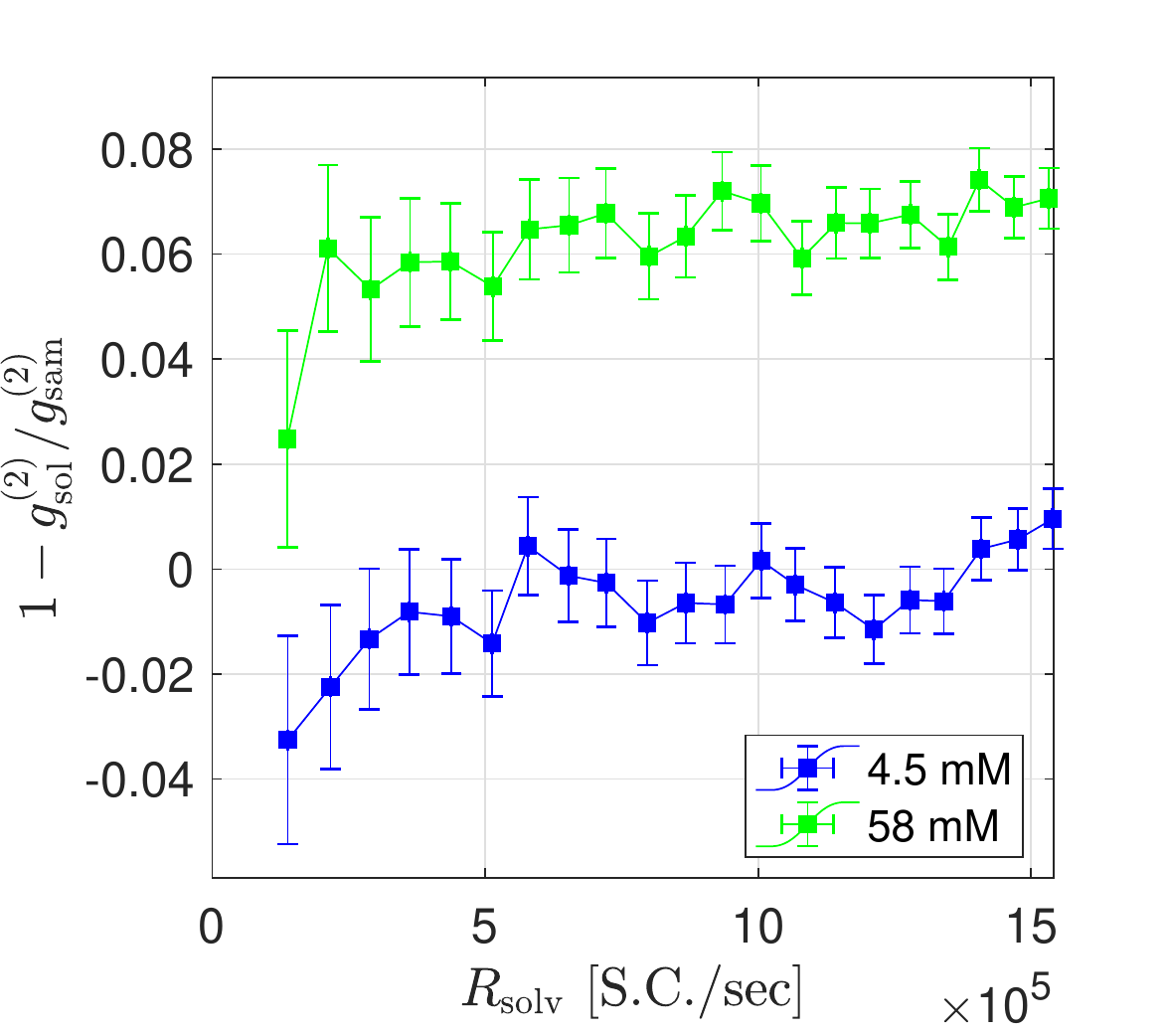}
    \includegraphics[width=0.49\textwidth]{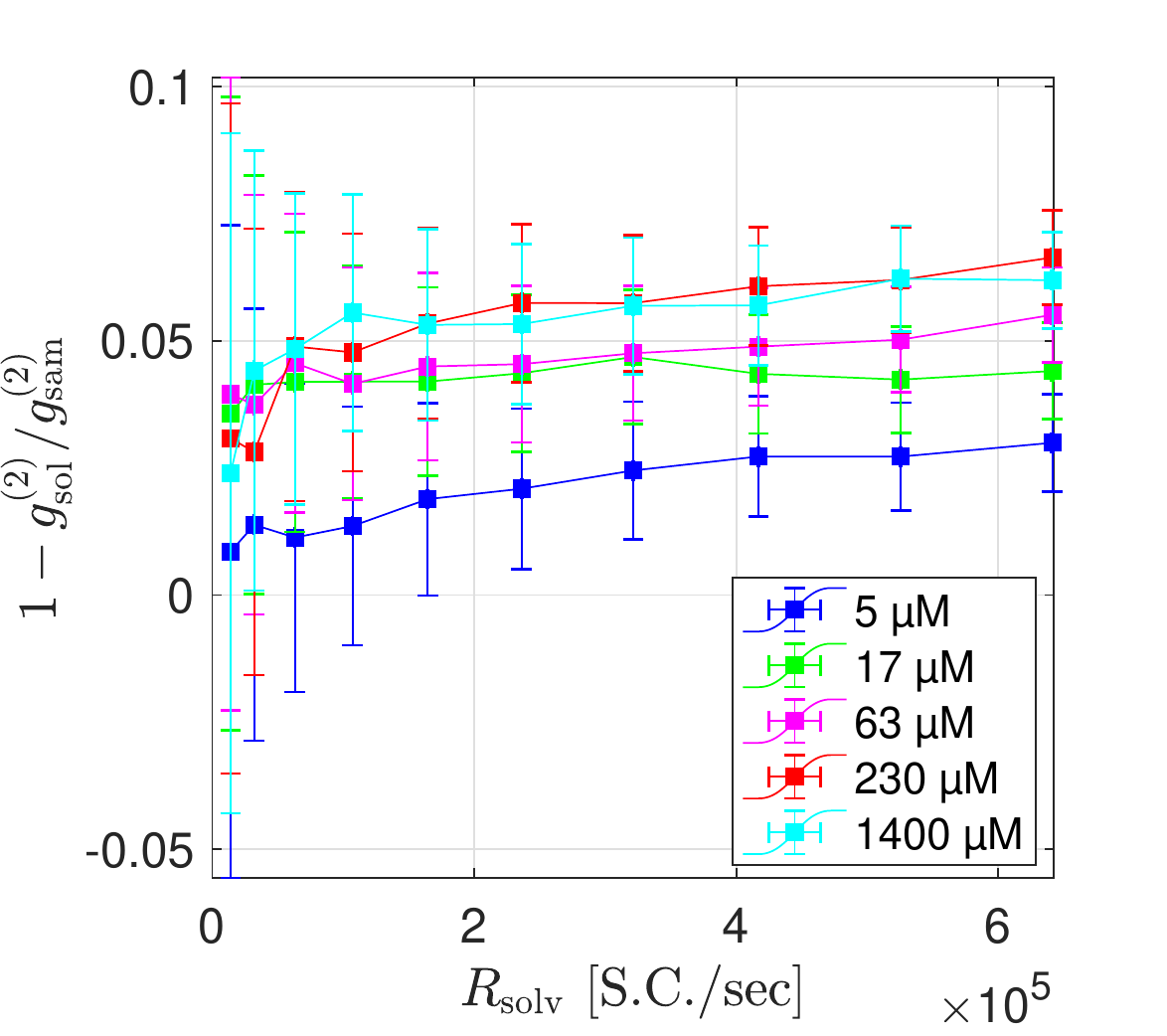}
    \caption{Experimental behaviour of $1- g^{(2)}_{0,\mathrm{solv}}/g^{(2)}_{0,\mathrm{sam}}$ as function of the solvent singles photon rate for different concentrations of RhB (left) and ZnTPP (right) in the collinear configuration.}
    \label{fig:fromg2&biphotons_rhb_colineal}
\end{figure}

We now set out to compare the results obtained through the standard scheme, with those obtained through our proposal based on second-order correlation functions.  With this purpose in mind, we now focus on the numerator of equation~(\ref{eq:sigma_g2}). Its behavior is shown in Figure~\ref{fig:fromg2&biphotons_rhb_colineal}, when plotted vs the input photon-pair flux (singles counts), where it can clearly be observed that such a value is approximately constant (beyond a specific flux level), indicating a linear relationship between
$\bar{R}^{(2)}_\mathrm{abs}$ and $\bar{R}^{(2)}_\mathrm{solv}$, as expected. In addition, the cross-sections obtained using the equation~(\ref{eq:sigma_g2}) are reported in the third column of the tables~\ref{Table: RhBCol} and \ref{Table: ZnTPPCol}.

Regardless of the definition used to calculate the ETPA cross-sections (i.e. relying on the standard scheme, or on our own proposal), the resulting values are within the same order of magnitude, i.e. $\sim10^{-20}$ for tens of mM of RhB, and $\sim10^{-18}$ for tens of $\mu$M of ZnTPP. Interestingly, for both types of solution, RhB and ZnTPP, the resulting cross-section values are larger than the lower bound established by equation~(\ref{eq:lowerbound}). While this suggests that the ETPA process may indeed be responsible for the observed signal, we have nevertheless designed a further experimental test.  In particular, in the next section we describe a modified experiment in which we introduce a temporal delay $\tau$ between the signal and idler photons, prior to reaching the ETPA sample.   The expected behavior is that for $\vert \tau \vert > T_\mathrm{e}$ where  $T_\mathrm{e}$ represents an SPDC photon pair characteristic time (also referred to as entanglement time), the ETPA process should be fully suppressed. \cite{Fei-1997,Saleh-1998,DLMontiel-2013}

Note that as mentioned above the estimation of the two-photon flux $R^{(2)}$ (equation \ref{eq:measbirate}) works irrespective of linear losses.  However, the ratio of $\tilde{R}^{(2)}_\mathrm{abs}$ to $\tilde{R}^{(2)}_\mathrm{solv}$ in equation (\ref{eq:biphotonratios}) involves either transitioning from solvent to solution, or from one concentration to another.  While this is done with utmost care, it is possible that the cuvette is slightly disturbed leading to possible deflection of the photon pair stream, thus explaining negative values of the ETPA signal in figure \ref{fig:fromg2&biphotons_rhb_colineal} (as well as in  \ref{fig:1-g2_both}, see below).

\subsubsection*{Non-collinear-SPDC configuration}

\begin{figure}[htp]
	\centering
	\includegraphics[width=0.9\textwidth]{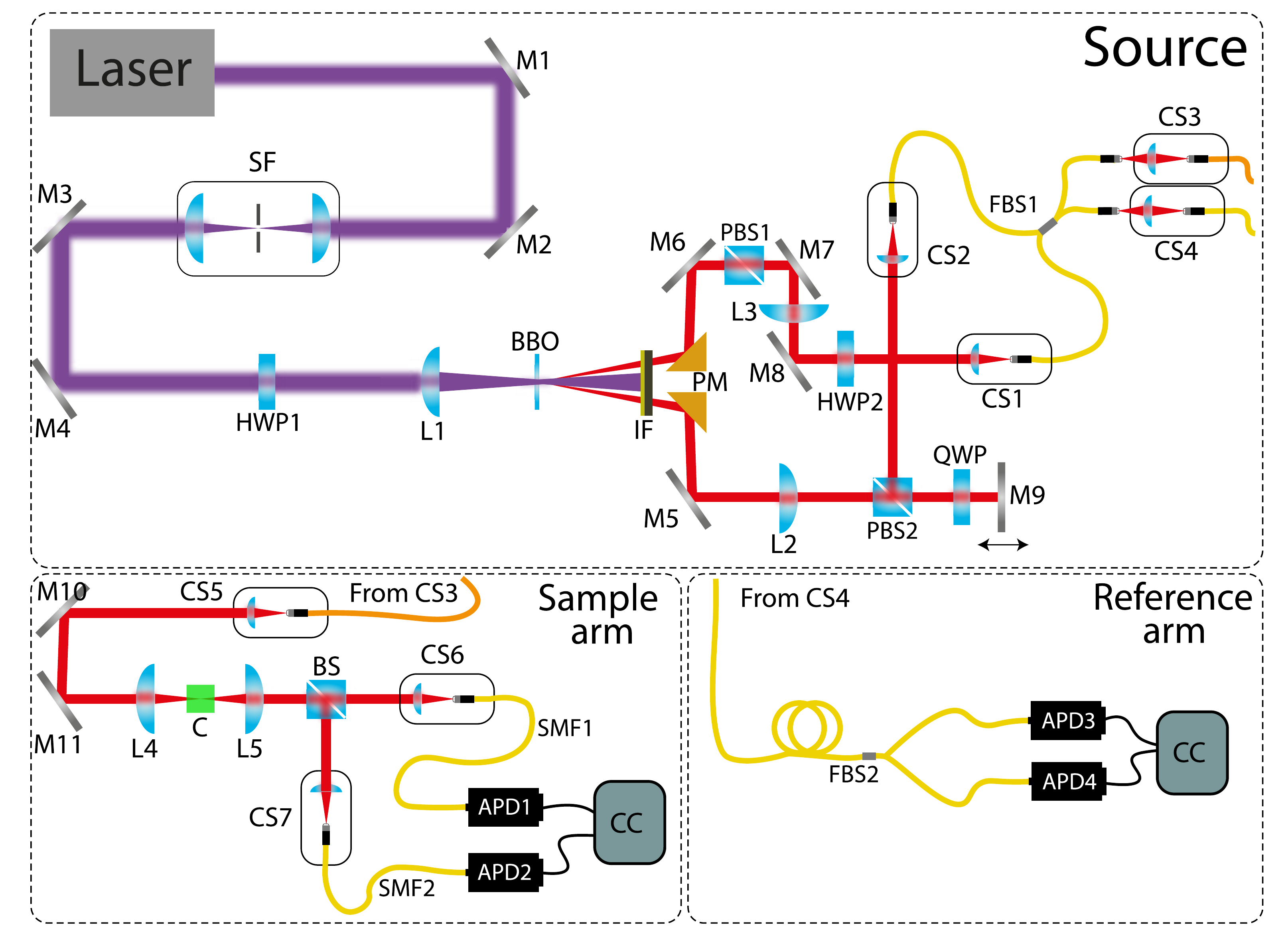}
	\caption{Entangled two-photon absorption with a non-collinear SPDC source. A spatially filtered (SF) CW-laser with controllable polarization (HWP1) pumps a BBO crystal to generate non-collinear photon pairs. The pairs are spectrally filtered (IF) and separated using prism mirrors (PM) into independent paths which are steered with mirrors (M5-M9), and collimated with lenses L2 and L3. The signal photon is delayed through a translatable mirror (M9). The polarizations of the signal and idler photons are both seet to vertical (PBS2, HWP2), to be coupled into a fiber beam splitter (FBS1). The resulting photon pair streams are sent to the sample and reference arms through a multimode fiber (orange), and a single-mode (yellow), respectively. The pairs in the sample arm pass through a quartz cuvette (C) containing the sample, and those that are transmitted are separated (BS) and detected with avalanche photo-diodes (APD1, APD2). In the reference arm,  the photon pairs are separated (FBS2) and detected with the avalanche photo-diodes (APD3, APD4).}
	\label{fig:SetupNocolineal}
\end{figure}

In order to investigate the behavior of the ETPA signal as a function of the temporal delay between the signal and idler photons, we now  switch to a non-collinear source configuration. It is worth pointing out that while in this configuration the photon pairs are emitted non-collinearly, their interaction with the sample remains collinear (see Figure~\ref{fig:SetupNocolineal}).  In this experimental design, we have relied on a Hong-Ou-Mandel (HOM) interferometer with the purpose of obtaining in each of the two output ports of the HOM fiber beamsplitter a stream of photon pairs, ensuring that both photons in each pair travel in identical spatial modes, with a controllable signal-idler temporal  delay.  While one stream of delayed photon pairs illuminates the ETPA sample, the other stream is employed as a reference, without sample.

The upper panel in Figure~\ref{fig:SetupNocolineal} shows the experimental setup for the non-collinear source configuration. As in the collinear configuration, we use a CW laser to pump a BBO crystal. The pump polarization is controlled using a half-wave plate (HWP1), which allows us to maximize the photon-pair flux. The BBO crystal is oriented so as to generate down-converted photon pairs at an angle of 2$^\circ$, with respect to the pump axis. Following the crystal, the pump beam is filtered out using interference filters (IF), while the SPDC cone is separated using two prism mirrors (PM). Lenses L2 and L3, both with focal length $f=100$ mm, are placed in each path so as to collimate the photons. The horizontally-polarized signal photon (in the lower path) can be temporally delayed with respect to the idler, by transmitting through a polarizing beam splitter (PBS2) and a quarter wave plate (QWP), so that upon reflection from a computer-controlled translatable mirror M9 the polarization is switched to vertical and the  photon is reflected at PBS2.

The position of mirror M9 used to reflect the signal-mode photons is controlled by a linear motor (Mercury C-863, Physik Instrumente -not shown in the diagram- with a minimum step of 50 nm), allowing us to introduce a deterministic signal-idler temporal delay. At the output of PBS2, the beam is coupled into one of the ports of a fiber beam-splitter (FBS1) with the help of a coupling system (CS2). To compensate the optical path, and select the horizontal polarization, the idler photon (top path) is transmitted through a polarizing beam splitter (PBS1), with the same characteristics as PBS2,  followed by a half-waveplate (HWP2) that switches its polarization to vertical and, finally, is injected into the second port of the fiber beam-splitter (FBS1). The signal and idler photons subsequently exhibit HOM interfere at FBS1. Note that by analyzing the width of the HOM dip, we find a correlation (entanglement) time $T_{e} = 0.20$ ps (FWHM), for a spectral width of 10 nm (see Appendix~\ref{App:ControllingDelay} for details). One of the output ports from FBS1 is coupled into a multi-mode fiber with coupling system CS3, and leads to the sample arm, shown in the lower-left panel of Fig.~\ref{fig:SetupNocolineal}, while the other output port of FBS1 is coupled into a single-mode fiber with coupling system CS4, and leads to the reference arm, shown in the lower-right panel of Fig.~\ref{fig:SetupNocolineal}.



In the sample arm, the light emerging from the multimode fiber output is collimated using coupling CS5 system. The photon-pair beam, reflected from mirrors M10 and M11, is focused on the sample using aspheric lens L4, with focal length $f=11$ mm. We place the beam waist in the center of the cuvette (C), with a spot radius of $w_{0} \approx 5$ $\mu$m. The light emerging from the cuvette (and collected by aspheric lens L5) is then split into two spatial modes using a beam splitter (BS). The light emerging from each of the beamsplitter output ports is then coupled into single-mode fibers (SMF1, SMF2) using the coupling systems CS6 and CS7. The signals are finally detected with avalanche photo-diodes (APD1, APD2), recording singles and coincidence counts using the time to digital converter (CC).

In the reference arm, the photon pair stream travels through a single mode fiber to reach a second fiber beamsplitter (FBS2).  The two output ports of FBS2 are directly connected, i.e. unaffected by the ETPA sample,  to the input ports of two single-photon detectors  (APD3 and APD4).  This reference arm constitutes a useful innovation, as it allows to compare the behavior a photon-pair stream transmitted through an ETPA sample and reaching a pair of single-photon detectors, with a fully equivalent photon-pair stream which never interacts with the ETPA sample.


Measurements of singles and coincidence counts as a function of the incoming photon flux are then obtained in both the sample and reference arms, and for both types of sample (RhB and ZnTPP).   We have carried out this collection of measurements  for two distinct delay settings: i) $\tau=0$, and ii) a particular delay value $\tau_0$, selected so that $\vert \tau_0 \vert \gg T_e$,  where $T_e$ is the photon pair characteristic time, also known as entanglement time.   Note that we expect to observe ETPA for the first delay setting ($\tau=0$), for which both photons in each pair are overlapped in time. Evidently, we expect that two-photon absorption is suppressed in the second delay setting ($\vert \tau_0 \vert \gg T_e$), as even the classical two-photon absorption process cannot take place at our extremely low photon flux ($\sim 10^5$ photon pairs/s at the maximum generation rate).

\subsubsection*{Results: Non-collinear SPDC configuration}

\begin{figure}[t!]
	\centering
	\includegraphics[width=0.49\textwidth]{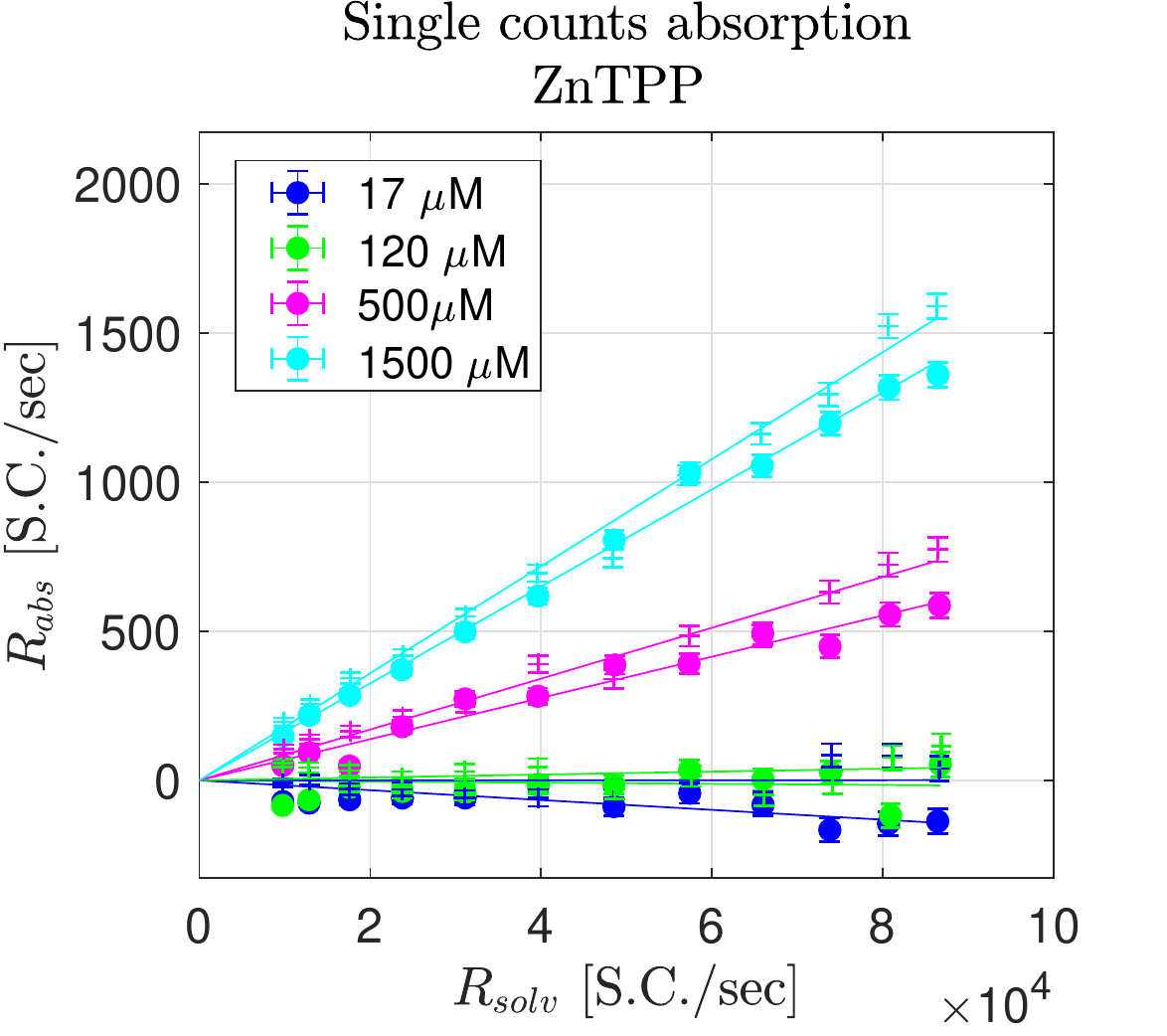}
	\includegraphics[width=0.49\textwidth]{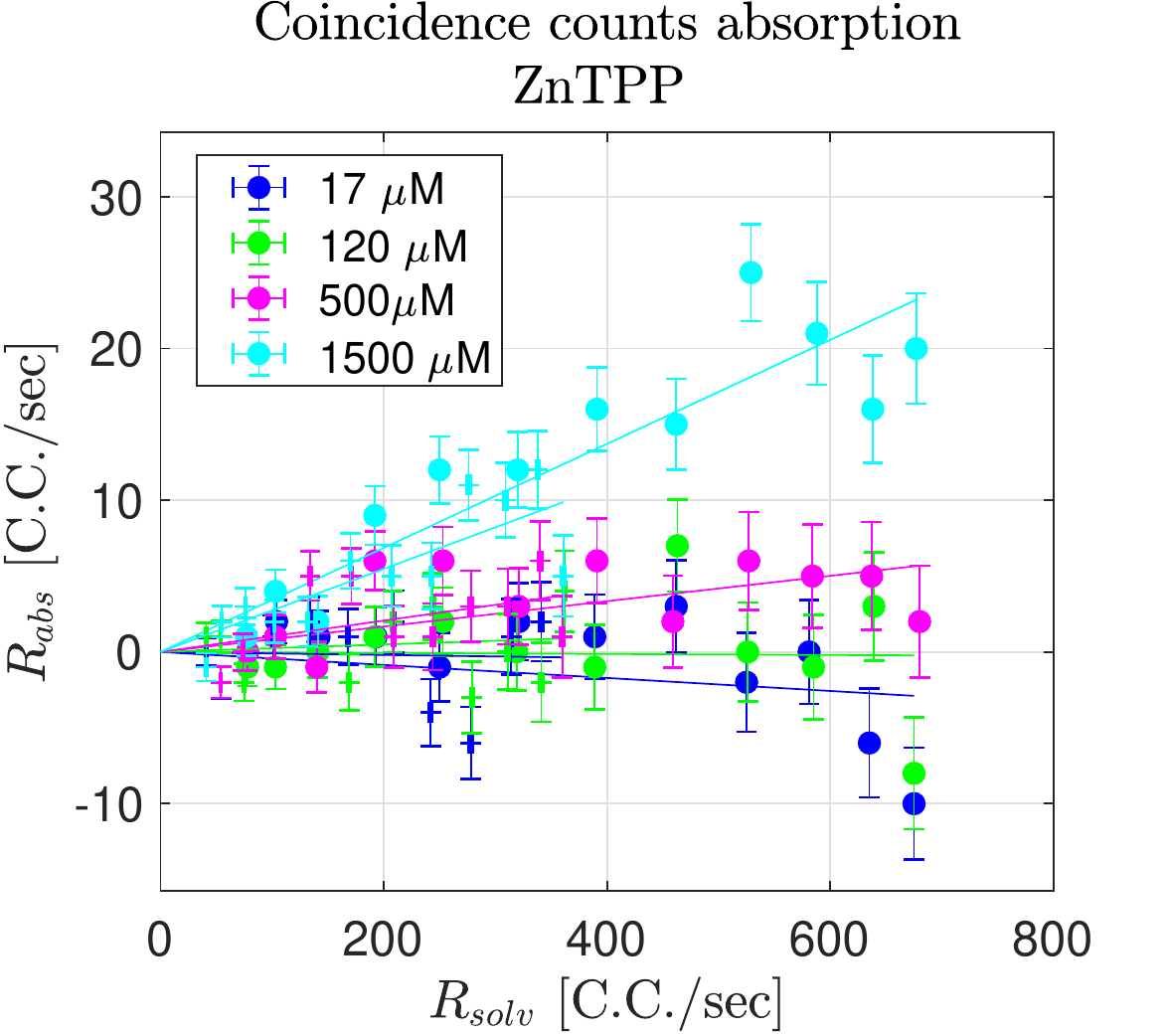}
	\includegraphics[width=0.49\textwidth]{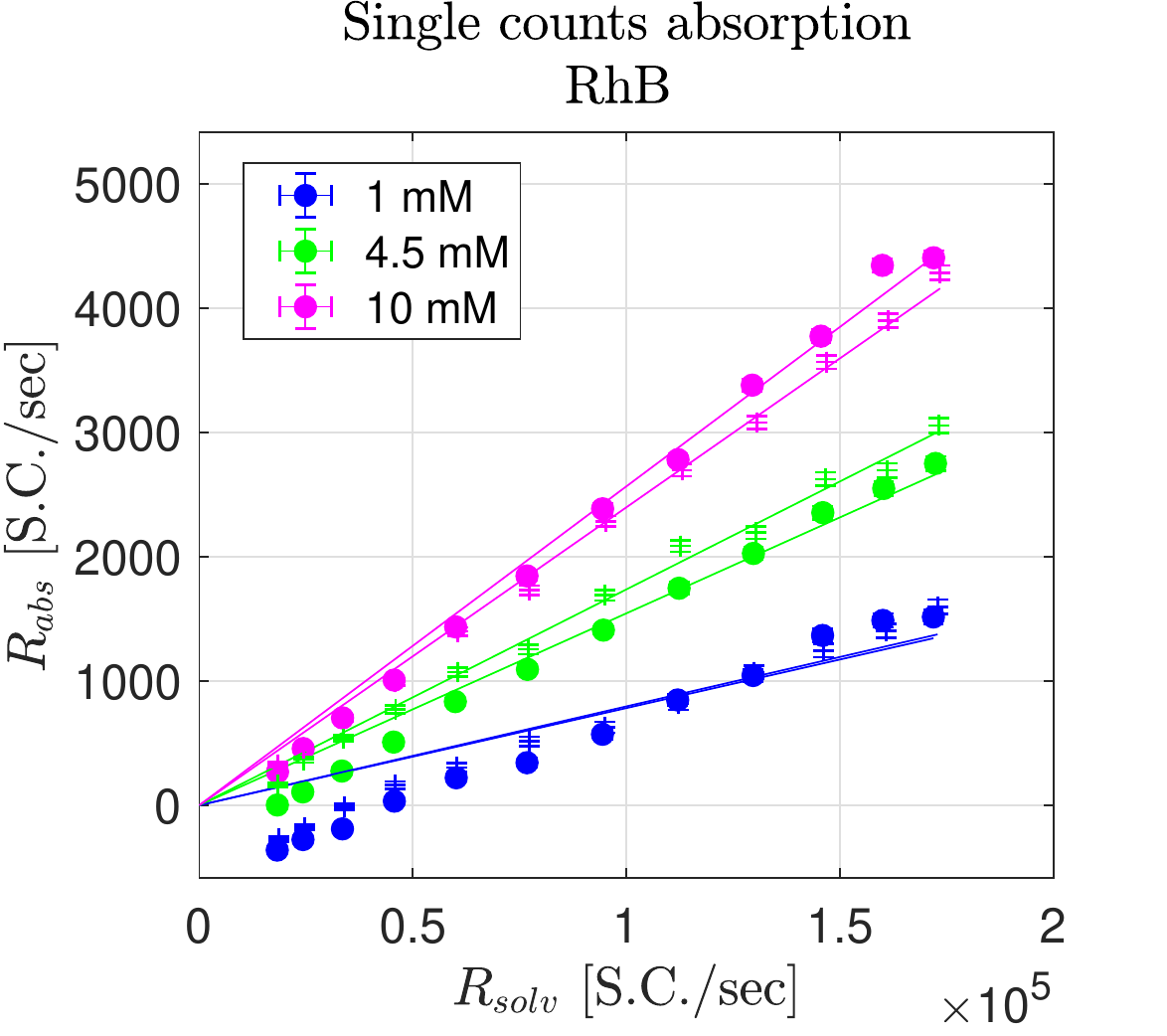}
	\includegraphics[width=0.49\textwidth]{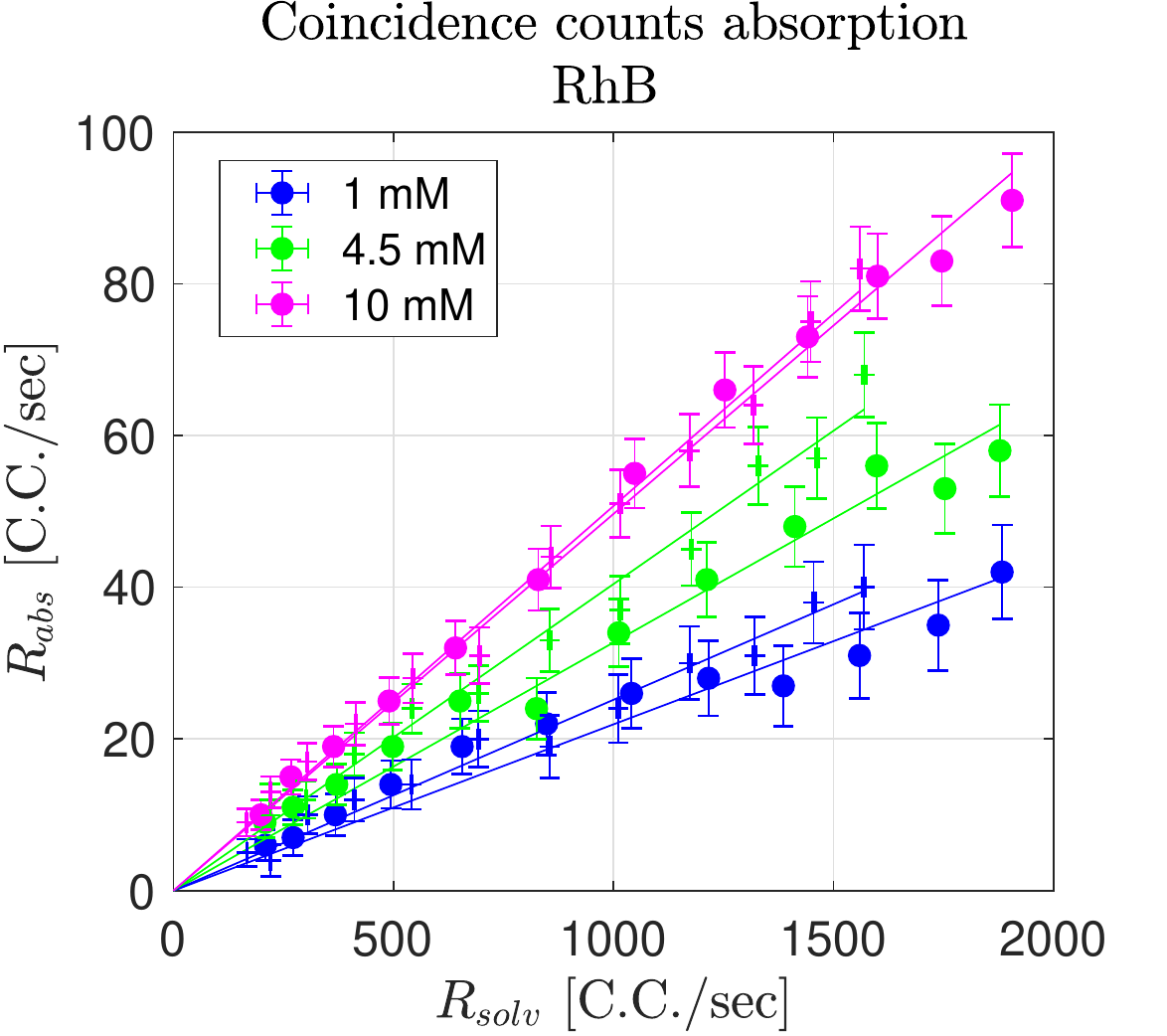}
	\caption{Absorption measured in single counts (left) and coincidence counts (right) for ZnTPP (upper row) and RhB (lower row) with (cross marker) and without delay (filled circles). Note that, for the sake of clarity, we have reduced the length of the error bars by a factor of $1/10$. Also note that, for the coincidence-count measurements, the maximum coincidence rate in the delayed setting is reduced with respect to the non-delayed configuration, see Appendix A for details.}
	\label{fig:AbsZnTPPCoincDentroyFuera}
\end{figure}

Figure~\ref{fig:AbsZnTPPCoincDentroyFuera} shows the experimental ETPA rates, measured by monitoring coincidence and singles counts for the delayed (cross symbols), and for the non-delayed (filled circles) experimental settings. For each of the two types of samples, different concentrations were tested: 17, 120, 500, and 1500 $\mu$M, for ZnTPP, and 1.0, 4.5, and 10.0 mM, for RhB. While the upper row corresponds to ZnTPP and the lower row corresponds to RhB, singles (coincidence) counts are shown in the left-hand (right-hand) columns.
Note that amongst the various experimental runs performed, we have selected the highest-quality ones which involve, for ZnTPP, the use of a $40$ nm-width spectral filter acting on the SPDC photon pairs, and an integration time of $60$s, while, for RhB, an $80$ nm-width filter, and a $180$s integration time.  The values for the non-zero delay $\tau_0$ used for the delayed experimental setting are $333$ fs and $667$ fs, for ZnTPP and RhB, respectively.

The error bars represent the Poissonian uncertainty, whereas the solid lines correspond to linear fits. As already mentioned, our expectation is to observe a non-zero absorption signal for the non-delayed experimental setting ($\tau=0$), and a vanishing absorption signal for the delayed setting $\tau=\tau_0$. However, this is in fact not what is observed experimentally. Our results indicate that the absorption signal has similar values, whether obtained from singles or coincidence counts, in  both the non-delayed and delayed settings.

It is worth noting that, because the ETPA process is not expected to occur for the delayed experimental setting (involving $\vert \tau \vert \gg T_e$), we would expect the slope of the $R_\mathrm{abs}$ vs $R_\mathrm{solv}$ measurement in the delayed setting to be zero for all of the concentration values. However, it becomes clear that, in figure~\ref{fig:AbsZnTPPCoincDentroyFuera}, different non-zero slopes are obtained for different concentrations.   This implies that the experiment might be susceptible to other effects, related to the concentration, such as attenuation and scattering due to molecular aggregation. Indeed, these processes hamper our ability to obtain the correct value of the ETPA cross-section solely from the $R_\mathrm{abs}$ vs $R_\mathrm{solv}$ slope.

\begin{figure}[t!]
	\centering
	\includegraphics[width=0.95\textwidth]{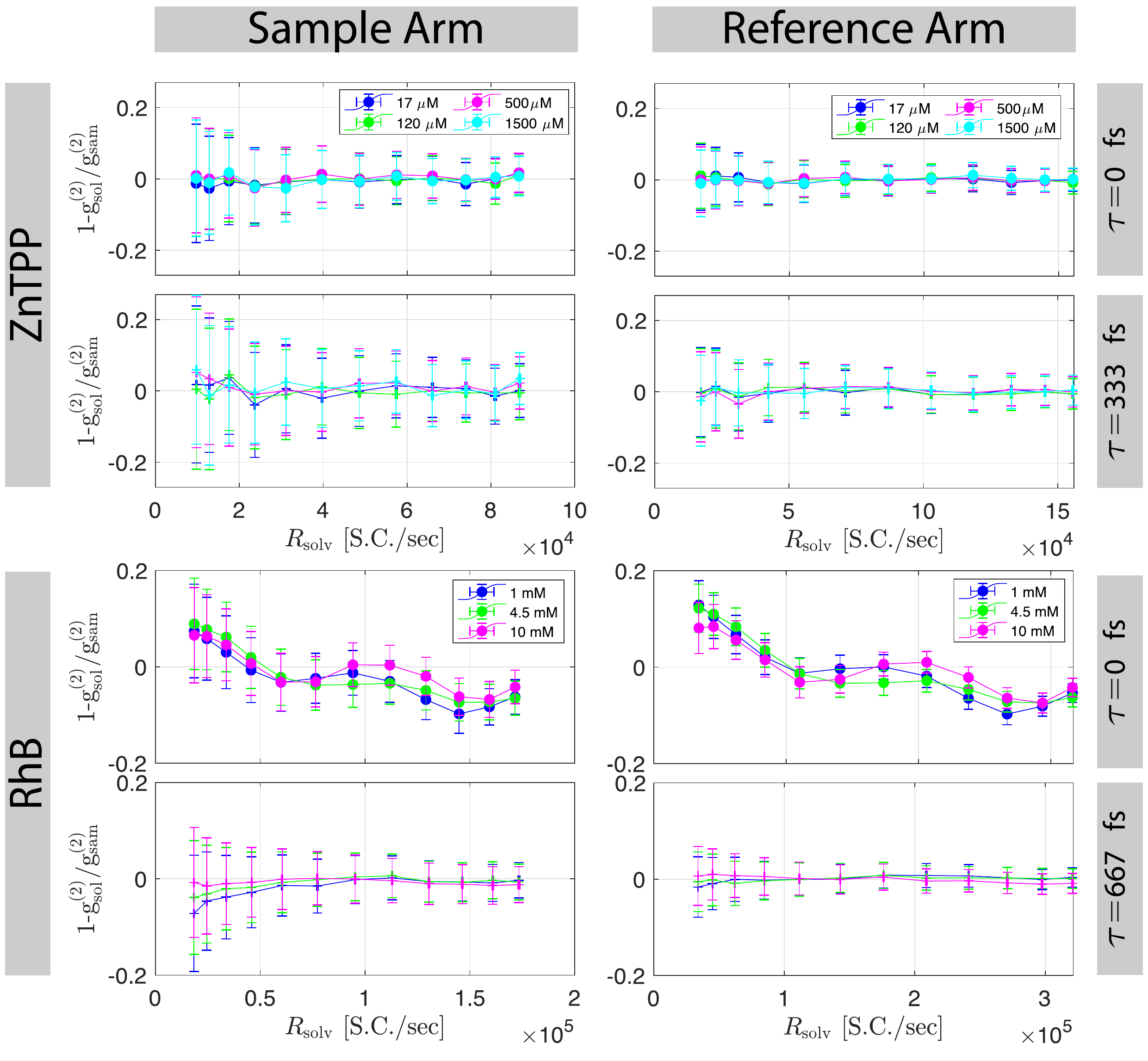}
	\caption{$1 -  g^{(2)}_{0,\mathrm{solv}}/g^{(2)}_{0,\mathrm{sam}}$ measured in the sample arm (left) and the reference arm (right) for ZnTPP (top two rows) and RhB (lower two rows).  Within each two-row block, the first row corresponds to the non-delayed setting, and the bottom row to the delayed setting.}
	\label{fig:1-g2_both}
\end{figure}

As in the case of the collinear source configuration, let us now shift from the standard scheme to our proposal based on second order correlation functions.   To this end in figure~\ref{fig:1-g2_both} we plot experimentally-obtained values for  $1 - g^{(2)}_{0,\mathrm{solv}}/g^{(2)}_{0,\mathrm{sam}}$ vs the photon pair flux $R_\mathrm{solv}$. Note that we have included data for, both, the non-delayed  and delayed  experimental settings.   While the left-hand column corresponds to the sample arm and the the right-hand column  to the reference arm, the first two rows correspond to ZnTPP, and the bottom two rows to RhB.  Within each block of two rows, the top row corresponds to the non-delayed $\tau=0$ setting, and the bottom row to the delayed setting ($\tau_0=333$ fs for ZnTPP and $\tau_0=667$ fs for RhB).

It is worth remarking that the photon pair flux $R_\mathrm{solv}$ axes have a different span of values for the sample and reference arms.  This is because in the case of the sample arm,  the photon pairs interact with optical elements in free space and are subsequently coupled into single mode fibers to be detected, while in the case of the reference arm,  propagation of the photon pair occurs entirely through optical fiber (see figure~\ref{fig:SetupNocolineal}). Data points for both arms can be mapped one to one since they were acquired concurrently as part of the same experimental run. The result is a horizontal scaling of the [$1 - g^{(2)}_{0,\mathrm{solv}}/g^{(2)}_{0,\mathrm{sam}}$ vs  $R_\mathrm{solv}$] experimental curves for the reference arm, as compared to the sample arm.

Because the reference arm contains no ETPA sample, we use its behavior as control (see Appendix~\ref{app:data_treatment}). In particular, since ETPA evidently does not occur in the \emph{reference} arm, if ETPA does indeed occur in the \emph{sample} arm, then the behaviors in the two arms must differ. In addition, as observed in previous experimental works we expect that the ETPA absorption rate should depend on the concentration,~\cite{villabona_calderon_2017} so that the experimental curves in figure~\ref{fig:1-g2_both} should exhibit a clear dependence on this parameter, as is the case in Figure~\ref{fig:AbsZnTPPCoincDentroyFuera}.    From an examination of figure~\ref{fig:1-g2_both} it becomes clear that
the sample and reference signals in fact show nearly identical behaviors (within statistical fluctuations). In addition, it is also clear from our experimental data that the [$1 - g^{(2)}_{0,\mathrm{solv}}/g^{(2)}_{0,\mathrm{sam}}$ vs  $R_\mathrm{solv}$] plots do not show a systematic dependence on the concentration; instead, they tend to fluctuate around a zero mean value. The fact that there is a clear dependence on the concentration in figure~\ref{fig:AbsZnTPPCoincDentroyFuera} can be explained by taking into account that in the standard scheme used to estimate the absorption signal, linear losses cannot be discriminated from the ETPA signal. In striking contrast, in our proposal based on second-order correlation functions, such linear losses are excluded.

Therefore, because figure~\ref{fig:1-g2_both} shows no dependence on the concentration, and because the sample and reference present similar behaviours, \emph{we conclude that there is no observable ETPA signal in our experiment}. We believe that the variations of the signal with respect to the concentration in figure~\ref{fig:AbsZnTPPCoincDentroyFuera} are due exclusively to linear loss processes (such as attenuation and scattering) which are impossible to discern in the standard approach.

\section{Conclusions}

We have proposed a new method for the estimation of entangled two-photon absorption (ETPA) cross section in a transmission-based experiment. Unlike the standard scheme based on the ratio of singles or coincidence counts obtained by switching the sample and the solvent, our method relying on the second order correlation functions, works irrespective of any linear loss mechanisms and/or any intrinsic dependence of the photon-pair flux on the temporal delay introduced between the signal and idler photons.   We have performed two ETPA experiments, one based on a collinear SPDC source in which we are able to vary the photon pair-flux and the sample concentration, and a second experiment in which we add a third variable, namely the signal-idler temporal delay.  Furthermore, we have presented a new experimental arrangement in which two separate measurement stages are introduced: the sample arm in which the photon pairs traverse an ETPA sample, and a reference arm without the sample. Given that equivalent photon-pair streams, obtained from the two output ports of a Hong-Ou-Mandel interferometer, are sent into each of these arms, the reference arm acts as a useful control to assess the behavior of the sample arm.

We find that our own measurements in the standard scheme, obtained in the first experiment, are compatible with those obtained by previous authors, with cross sections fulfilling the bound established in equation~(\ref{eq:lowerbound}).  We nevertheless also find that in our second experiment: i) the ETPA signal is not suppressed for temporal delays $\tau$ greater than the photon-pair characteristic time $T_e$, i.e. $\vert \tau \vert \gg T_e$, and ii) the ETPA signal obtained through our method (based on second order correlation functions) is not appreciably different for the sample and reference arms, and does not have a clear dependence on the sample concentration.  Having considered these findings, 
it is fair to conclude that the transmission-based methods that are currently employed to explore ETPA are not entirely reliable. Our work and that presented by Li,~\cite{li_SL_2020} in which only a 40 times enhancement of the absorption signal was reported (despite using large fluxes of strongly correlated photons), suggest that the available flux of entangled photons in our work and several others with similar photon-pair flux levels, is insufficient to produce an ETPA signal that is strong enough to be distinguished from other effects such as linear absorption and scattering.  Alternatively, the cross-sections of these molecules could be  inaccessible through a transmission-based method, which is in agreement with the work presented by Parzuchowski \emph{et al}.~\cite{Parzuchowski-2020}
We believe that our work will shed light on the current debate regarding the viability of the transmission-based scheme for ETPA in organic compounds.

\begin{acknowledgments}

R.J.L.M. thankfully acknowledges financial support by CONACyT under the project CB-2016-01/284372 and by DGAPA-UNAM under the project UNAM-PAPIIT IN102920.  A.U. acknowledges support from DGAPA-UNAM through grant UNAM-PAPIIT IN103521, and CONACyT through "Ciencia de Frontera" grant 217559.

\end{acknowledgments}

\appendix

\section{Controlling the temporal delay}
\label{App:ControllingDelay}

We have used Hong-Ou-Mandel (HOM) interference~\cite{HOM-1987} so as to obtain at the two HOM beamsplitter (FBS1) output ports two equivalent streams of temporal-delayed photon pairs, with the signal idler guaranteed to be in the same spatial mode.

\begin{figure}
	\centering
	\includegraphics[width=0.49\textwidth]{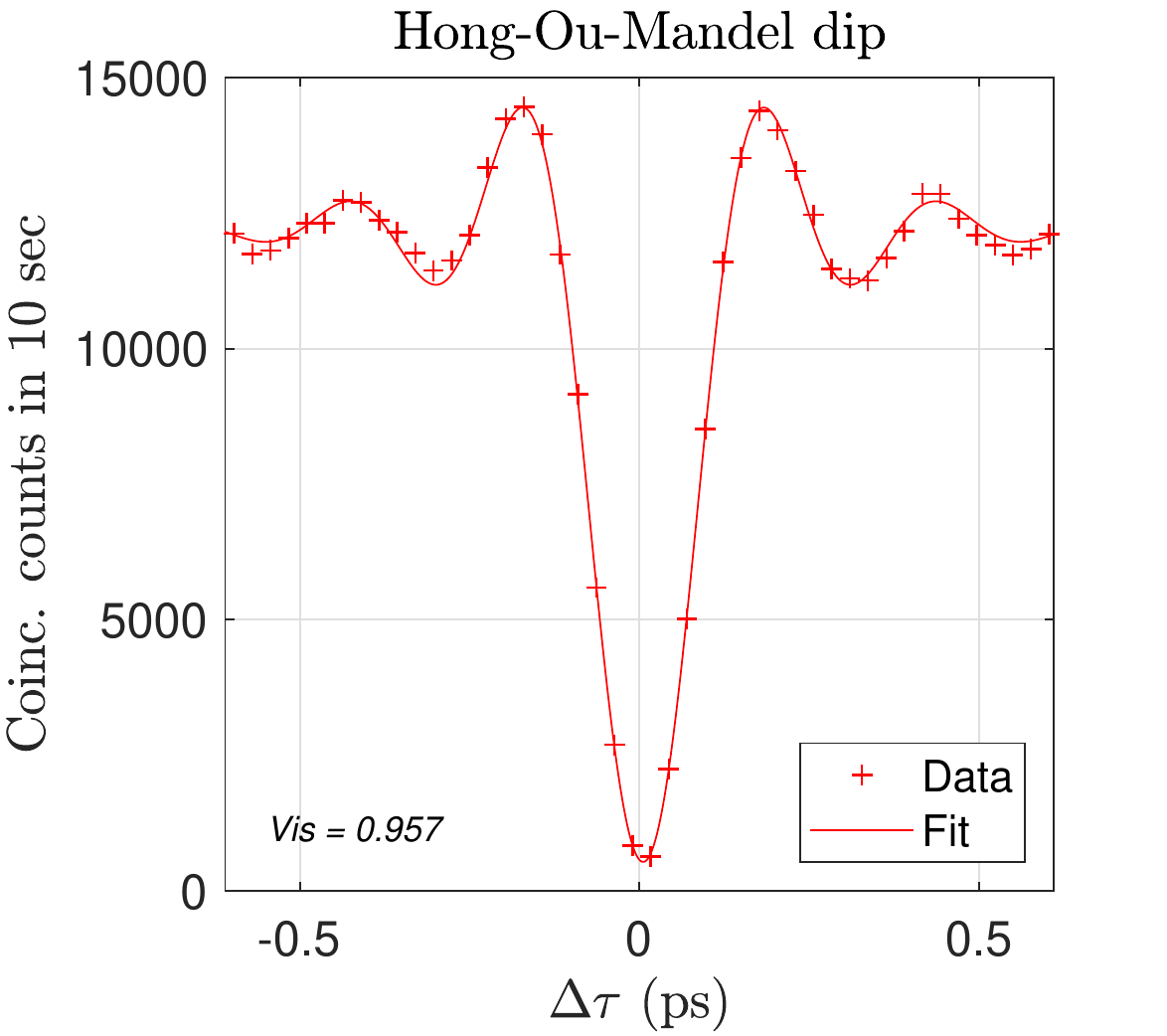}
	\includegraphics[width=0.49\textwidth]{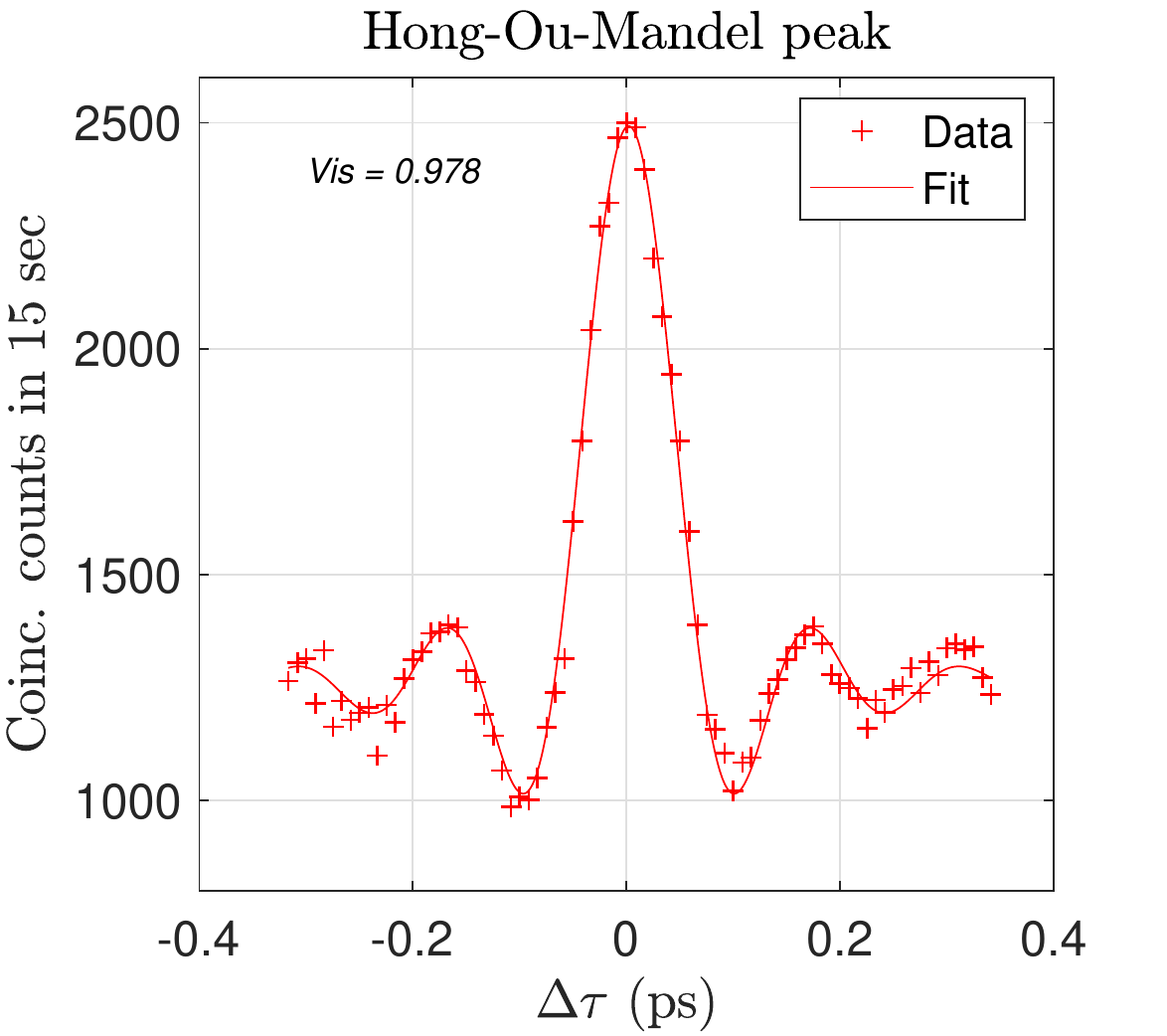}
	\caption{HOM interference. (left) Dip measured, with a 10 nm (FWHM) interference filter, (Right) bunching peak measured, with a 40 nm (FWHM) interference filter.}
	\label{fig:HOMdippeak}
\end{figure}

A standard HOM measurement is first obtained by removing the beamsplitters in the sample and reference arms (BS and FBS2), placing a single detector in each of the two arms, and monitoring coincidence counts in these two detectors as function of the temporal delay (obtained by translating mirror M9).  We characterize the coherence time of the SPDC pair through the width of the resulting HOM dip, shown in the left-hand panel of Figure~\ref{fig:HOMdippeak}, for which an integration time per experimental point of 10 seconds was used.  The solid line represents the best fit according to the model $a + d  \mathrm{sinc}[c_{1}(x-b)]\exp\left[- \left(\frac{x-b}{c_2}\right)\right]^2 $. Note that the resulting  HOM visibility is $V=0.957$  and the HOM dip full width at half maximum, or photon-pair correlation time, is $\sim 0.2 ps$ (for a filter width of $10$nm acting on the SPDC photon pairs).


With the full setup shown in Fig. \ref{fig:SetupNocolineal} in place, because a photon pair stream travels through each of the arms, a bunching peak, instead of a dip, is expected when monitoring the coincidence count rate in any of the two arms vs the temporal delay.   In the right-hand panel of Fig.~\ref{fig:HOMdippeak} we show such a bunching peak for the reference arm, obtained with a $15$ second integration time per experimental point, with the SPDC photon pairs transmitted through a $40$ nm width spectral filter.  It is clear from this result that there is an intrinsic dependence of the photon-pair flux on the temporal delay which is unrelated to ETPA (specifically, with the coincidence rate at $\tau=0$ being $2$ times larger than that at $\vert \tau \vert \gg T_e$).  In an ETPA experiment in which an intra photon-pair delay is used, care must be taken not to confuse the expected ETPA delay dependence of the coincidence rate with the intrinsic delay-dependence shown here.   Our proposal based on second-order correlation functions has the crucial advantage of working irrespective of this intrinsic delay dependence of the photon-pair flux.

\section{Reference-based error estimation}
\label{app:data_treatment}
As mentioned in the main text, the use of a reference arm in our non-collinear SPDC configuration experiment constitutes a useful control for the behavior of the sample arm. In particular, the reference arm can be employed so as to correct the sample-arm signal for external fluctuations, e.g. pump laser variations, temperature, or mechanical vibrations affecting the HOM interferometer. To account for these possible errors, a correction factor for each measurement in the reference arm was introduced. This factor was computed as the mean of the quotient between the signal measured (in the reference arm) in the presence of the solvent and that corresponding to each of the different concentrations. The correction factor was then applied to the sample arm for all data, whether displayed according to the standard scheme or according to our proposal based on second order correlation functions.

\section*{Data availability}
The data that support the findings of this study are available from the corresponding author upon reasonable request.
\bibliography{ETPA_biblio}

\end{document}